\documentclass[twocolumn]{aastex631}

\usepackage{booktabs}
\usepackage{amsmath, multirow}
\usepackage{rotating}
\usepackage{longtable}

\usepackage[normalem]{ulem} 
\usepackage{soul,xcolor} 
\setstcolor{red} 

\def\HI{{\rm H\,{\textsc{\romannumeral 1}}}}
\def\HII{{\rm H\,{\textsc{\romannumeral 2}}}}
\def\HIMF{{\rm H\,{\textsc{\romannumeral 1}}MF}}
\def\HIWF{{\rm H\,{\textsc{\romannumeral 1}}WF}}

\graphicspath{{./}{figures/}}

\begin{document}

\title{FAST Ultra-Deep Survey (FUDS): Data Release for FUDS0}

\correspondingauthor{Bo Peng}
\email{pb@nao.cas.cn}
\correspondingauthor{Lister Staveley-Smith}
\email{lister.staveley-smith@uwa.edu.au}

\author[0000-0001-6642-8307]{Hongwei Xi}
\affiliation{National Astronomical Observatories, Chinese Academy of Sciences\\
20A Datun Road, Chaoyang District, Beijing 100101, China}

\author[0000-0001-6956-6553]{Bo Peng}
\affiliation{National Astronomical Observatories, Chinese Academy of Sciences\\
20A Datun Road, Chaoyang District, Beijing 100101, China}
\affiliation{Department of Astronomy and Institute of Interdisplinary Studies, Hunan Normal University\\
Changsha, Hunan 410081, China}

\author[0000-0002-8057-0294]{Lister Staveley-Smith}
\affiliation{International Centre for Radio Astronomy Research (ICRAR), University of Western Australia,\\ 
35 Stirling Hwy, Crawley, WA 6009, Australia}
\affiliation{ARC Centre of Excellence for All Sky Astrophysics in 3 Dimensions (ASTRO 3D), Australia}

\author[0000-0002-0196-5248]{Bi-Qing For}
\affiliation{International Centre for Radio Astronomy Research (ICRAR), University of Western Australia,\\ 
35 Stirling Hwy, Crawley, WA 6009, Australia}
\affiliation{ARC Centre of Excellence for All Sky Astrophysics in 3 Dimensions (ASTRO 3D), Australia}

\author[0000-0002-1311-8839]{Bin Liu}
\affiliation{National Astronomical Observatories, Chinese Academy of Sciences\\
20A Datun Road, Chaoyang District, Beijing 100101, China}
\affiliation{Department of Astronomy and Institute of Interdisplinary Studies, Hunan Normal University\\
Changsha, Hunan 410081, China}

\author[0000-0002-7550-0187]{Dejian Ding}
\affiliation{National Astronomical Observatories, Chinese Academy of Sciences\\
20A Datun Road, Chaoyang District, Beijing 100101, China}
\affiliation{School of Astronomy and Space Science, University of Chinese Academy of Sciences\\
No.1 Yanqihu East Road, Huairou District, Beijing, 101408, China}



\begin{abstract}


    We have used the Five-hundred-meter Aperture Spherical radio Telescope (FAST) to make a blind ultra-deep survey for neutral hydrogen (\HI). We present the complete results from the first of six fields (FUDS0). This observation of 95 hours allowed us to achieve a high sensitivity ($\sim 50~\mu$Jy~beam$^{-1}$) and a high frequency resolution (22.9~kHz) over an area of 0.72~deg$^2$. We detected 128 galaxies in \HI\ distributed over the redshift range of $0<z<0.4$ with \HI\ masses in the range of $6.67 \leq \log(M_{\rm \HI}/h_{70}^{-2} \rm M_\odot) \leq 10.92$, and three faint high-velocity clouds (HVCs) with peak column density of $N_{\rm HI} \leq 3.1 \times 10^{17}$~cm$^{-2}$. Of the galaxies, 95 are new detections and six have $z > 0.38$, where no unlensed \HI\ emission has previously been directly detected. Estimates of completeness and reliability are presented for the catalog. Consistency of continuum and \HI\ flux estimates with NVSS and AUDS, respectively, confirms the accuracy of calibration method and data reduction pipeline developed for the full FUDS survey.

\end{abstract}



\section{Introduction}\label{Sct_01}
    
    Hydrogen is the most abundant element in our Universe. Three forms are found in galaxies: neutral atomic (\HI), molecular (H$_2$) and ionized (\HII). \HI\ is the most common phase in the interstellar medium (ISM) of late-type galaxies. Collapsing \HI\ clouds result in the formation of molecular clouds by the conversion of \HI\ into H$_2$, mainly on the surface of dust grains \citep{1963ApJ...138..393G}. Molecular clouds are the birthplaces of most stars. Accretion of gas from satellites and the intergalactic medium (IGM) replenishes \HI\ gas, which helps sustain ongoing star formation in galaxies \citep{2008A&ARv..15..189S, 2012ARA&A..50..531K}. The radiation, mechanical and thermal energy injected by massive stars into interstellar medium (ISM) also helps maintain the ionized phase, \HII, which cools over time resulting in a replenishment of the neutral \HI\ phase. In the so-called baryonic cycle, \HI\ serves as fuel for stars, and indirectly regulates star formation. Studies of \HI\ in the ISM and its accretion onto galaxies is therefore pivotal for understanding the formation and evolution of galaxies.
    
    Large samples of \HI\ measurements for galaxies have historically been difficult to obtain due to the difficulty of measuring and mapping radiation from the weak 21-cm line \citep{1997ApJ...490..173Z, 1998ApJS..119..159S}. However, with the emergence of multi-beam receivers \citep[e.g.][]{1996PASA...13..243S}, the survey speed of many of large single-dish radio telescopes has been increased by an order of magnitude. The sky coverage of \HI\ surveys and the number of \HI\ detections of galaxies has increased in a commensurate manner. \citet{2003MNRAS.342..738L} conducted the \HI\ Jodrell All-Sky Survey (HIJASS) by employing a 4-beam receiver installed on the Lovell telescope to map an area of 1775 deg$^2$ in the northern sky ($\delta > +22^\circ$), resulting in 396 \HI-detected galaxies. Meanwhile, around 21,341 deg$^2$ in the southern sky ($\delta < +2$) was surveyed by Parkes telescope with equipped 13-beam receiver (\HI\ Parkes  All Sky Survey (HIPASS); \citealp{2001MNRAS.322..486B, 2004MNRAS.350.1195M}). HIPASS detected 4315 \HI\ galaxies. Another 1002 \HI\ galaxies were detected in an extended survey in the northern sky $+2\arcdeg < \delta < +25\fdg5$ \citep{2006MNRAS.371.1855W}. Based on the HIPASS catalogue, \cite{2005MNRAS.359L..30Z} derived the first accurate \HI\ mass function.
    
    The even larger aperture of the Arecibo radio telescope later equipped with the Arecibo $L$-band Feed Array (ALFA), led to further improvements in survey speed, sensitivity and resolution \citep{2005AJ....130.2598G}. The Arecibo Legacy Fast ALFA Survey (ALFALFA), which mapped 7000 deg$^2$ in the footprint of the Sloan Digital Sky Survey (SDSS), had a sensitivity of 2.4 mJy~beam$^{-1}$ and detected more than 30,000 \HI\ galaxies. By using this larger sample, improvements in statistical analyses of the \HI\ velocity width function (\HIWF) and the \HI\ mass function (\HIMF), and  dependence of environment, have been made \citep{2010ApJ...723.1359M, 2014MNRAS.444.3559M, 2016MNRAS.457.4393J, 2018MNRAS.477....2J, 2022MNRAS.509.3268O}. 
    
    The Commensal Radio Astronomy FAST survey (CRAFTS, \citealp{2018IMMag..19..112L, 2019SCPMA..6259506Z}) and FAST all sky \HI\ survey (FASHI, \citealp{2024SCPMA..6719511Z}), which are being carried out by using a 19-beam receiver installed on FAST telescope, are providing even larger samples in the northern sky for analysis. In the southern sky, the Widefield ASKAP $L$-band Legacy All-sky Blind surveY (WALLABY; \citealp{2020Ap&SS.365..118K}), which is carried out by the Australian SKA Pathfinder (ASKAP), is utilising a state-of-the-art phased array feed to detect and make high-resolution images of over $10^5$ galaxies.
    
    Due to limited sensitivity, wide area surveys such as the above have been confined to the local Universe and cannot provide any evolutionary information on galaxies and their \HI\ content. On the other hand, single-dish pencil beam surveys such as the Arecibo Ultra-Deep Survey (AUDS) have utilized the advantages of both the multi-beam receiver and a large illuminated aperture to blindly search for extragalactic \HI\ over small areas ($\sim 1.35$ deg$^2$) with a high sensitivity ($\sim$ 75 $\mu$Jy~beam$^{-1}$) \citep{2011ApJ...727...40F, 2015MNRAS.452.3726H, 2021MNRAS.501.4550X}. Based on the catalogue from AUDS 100\% data, \citet{2021MNRAS.501.4550X} presented  evidence for an evolutionary trend in the ``knee'' mass of the \HIMF. However, the small AUDS sample (247 \HI\ galaxies) only gives results of low significance. 
    
    Other recent deep surveys have been carried out by interferometers over wider redshift ranges in order to explore the evolution of the \HIMF\ and the \HIWF. In the northern hemisphere, the Blind Ultra Deep \HI\ Environmental Survey (BUDHIES; \citealp{2007ApJ...668L...9V}) has explored the evolution and environmental dependence of galaxy \HI\ content based on \HI\ galaxies detected in two distinct clusters at $z \simeq 0.2$. The Cosmos HI Large Extragalactic Survey (CHILES; \citealp{2016ApJ...824L...1F}) used Jansky Very Large Array (VLA) to survey 0.5 deg$^2$ COSMOS field at $z<0.45$ to a sensitivity of 50~$\mu$Jy~beam$^{-1}$. In the southern hemisphere, the Deep Investigation of Neutral Origins (DINGO; \citealp{2009PRA...........M}) will employ ASKAP to map 60 deg$^2$ within the Galaxy And Mass Assembly (GAMA) footprint up to $z=0.43$ with similar sensitivity (40~$\mu$Jy~beam$^{-1}$). The MeerKAT telescope will be employed to map 2~deg$^2$ area with sensitivity of 68 nJy to search for \HI\ galaxies up to $z=1.4$ in Looking At the Distant Universe with the MeerKAT Array survey (LADUMA; \citealp{2016mks..confE...4B, 2018AAS...23123107B}).

    The FAST telescope, with its large illuminated diameter ($\sim$ 300 m), multi-beam receiver system (19 beams) and wide band backend (1.0 -- 1.5 GHz), is capable of extending the exploration of \HI\ in galaxies to higher redshifts than Arecibo -- up to $z \sim 0.4$ with a reasonable integration time. In \citet{2022PASA...39...19X}, we presented the science goals, calibration techniques and preliminary results from a new blind \HI\ survey, the FAST Ultra-Deep Survey. This survey will map six 0.72 deg$^2$ fields with a high sensitivity ($\sim$ 50 $\mu$Jy~beam$^{-1}$), which greatly extends the sky coverage, sensitivity and redshift range of AUDS. The observations for the first area (FUDS0) are now completed. In this paper, we describe the FUDS0 observations and data reduction in Section \ref{Sct_02} and \ref{Sct_03}. The source finding methods and the catalogue are presented in Section \ref{Sct_04} and Section \ref{Sct_05}, respectively. The completeness and reliability of our FUDS0 catalogue are discussed in Section \ref{Sct_06} and \ref{Sct_07}. In Section \ref{Sct_08}, we compare the properties of FUDS0 galaxies with previous AUDS survey. We summarize our findings in Section \ref{Sct_09}. Throughout the paper, we use the flat universe model with parameters of $H_0=70~h_{70}$ km~s$^{-1}$ Mpc$^{-1}$, $\Omega_{\rm M}=0.3$, and $\Omega_\Lambda=0.7$.
    
\section{Observation}\label{Sct_02}

\begin{figure}
        \begin{center}
            \includegraphics[width=\columnwidth]{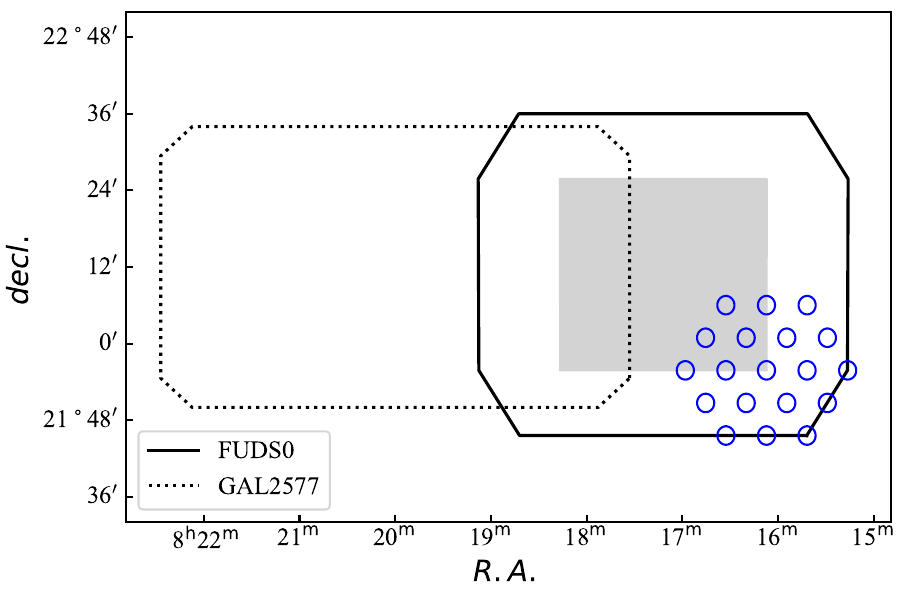}
            \caption{The survey areas of the FUDS0 and the AUDS GAL2577 field are shown with the dotted and solid black lines, respectively. The grey area shows the central high-sensitivity area of FUDS0. The FAST 19-beam receiver footprint is shown with the blue circles.}\label{Fig_01}
        \end{center}
    \end{figure}

    The FUDS survey consists of six target fields. The selection criteria, positions of the fields, observing and calibration methodology are described in \citet{2022PASA...39...19X}. Here we only give a brief summary of the FUDS0 observations. The FUDS0 field covers 0.72 deg$^{2}$, and was chosen to overlap with the GAL2577 field of the Arecibo Ultra-Deep Survey (AUDS, \citealp{2011ApJ...727...40F, 2015MNRAS.452.3726H, 2021MNRAS.501.4550X}) which has a similar sensitivity ($\sim 75~\mu$Jy~beam$^{-1}$) but a smaller redshift coverage ($z<0.16$). Comparison between our results and previous work is useful in verifying calibration and data reduction methods. The overlap of the fields \citep{2022PASA...39...19X} is shown in Figure \ref{Fig_01}. We selected the FUDS0 field and overlap region by excluding regions with strong continuum sources. In the NRAO VLA Sky Survey catalogue (NVSS, \citealp{1998AJ....115.1693C}), there are four continuum sources with flux density greater than 50~mJy in or close to FUDS0 field, with the strongest source is located at $R.A.$(J2000)=08:17:35.1, $decl.$(J2000)=+22:37:12 with flux density of 1.28\,Jy.
    
    The observations were taken between 2019 Aug 25 and 2020 May 22, which included part of the FAST commissioning phase. The FUDS0 field was scanned in both the $R.A.$ and $decl.$ directions with the 19-beam receiver in on-the-fly (OTF) mode \citep{2022PASA...39...19X}. The 500 MHz bandwidth (1.0 -- 1.5 GHz) was split into 65,536 channels, which gives a frequency resolution of 7.63 kHz (with equivalent velocity resolution 1.61 km~s$^{-1}$ at $z=0$). The spectra were recorded with a 1 s integration time. The total observation time was 129 hrs, consisting of 15 hrs on the calibrator, 95 hours on FUDS0, and 19 hours of overhead time (10 minutes are required for FAST to change its observing target).
    
\section{Data reduction}\label{Sct_03}

    The data reduction method is detailed in \citet{2022PASA...39...19X}. The final cube covers 1 deg$\times$1 deg, as shown in  Figure \ref{Fig_01}, with a pixel size of 1 arcmin $\times$ 1 arcmin. The gridded beam size is 3.27 arcmin at $z=0$ and its frequency dependence is given in \citet{2022PASA...39...19X}. The frequency spacing is 7.63 kHz but the resolution was changed to 22.9 kHz (4.83 km~s$^{-1}$ at $z=0$) by Hanning smoothing. Since the field is not uniformly sampled, the noise rises from the center to the edge, with the lowest noise being about 50 $\mu$Jy~beam$^{-1}$, measured by calculating the root mean square ($RMS$) in 186 channels (300 km~s$^{-1}$ or 1.4 MHz) of the Hanning-smoothed cube at $z=0$ in the frequency range free of radio frequency interference (RFI).

\begin{figure}
    	\begin{center}
    		\includegraphics[width=\columnwidth]{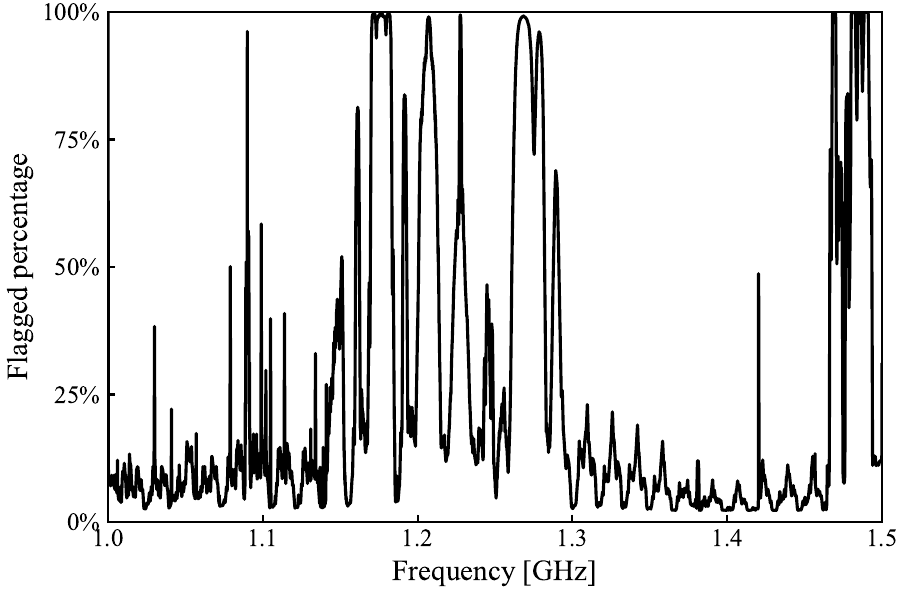}
    		\caption{The percentage of data flagged as a function of frequency. Virtually all flagging is due to the presence of RFI.}\label{Fig_02}
    	\end{center}
    \end{figure}

    We use the flagging procedures in \citet{2022PASA...39...19X}. The flagged fraction at different frequencies is shown in Figure \ref{Fig_02}. Most flagging is associated with external RFI, e.g. radar ($\sim$1.09 GHz), global navigation satellite system (GNSS, 1.15 -- 1.30 GHz), and the geostationary AsiaStar satellite ($\sim$1.48 GHz). We also exclude the \HI\ emission line from Milky Way for this analysis. FAST also suffered from strong internal compressor RFI prior to 2021 July, which severely impacted the FUDS0 observations, causing a quasi-periodic change in the flagged fraction in the frequency range 1.30 -- 1.46 GHz (see Figure~\ref{Fig_02}). The total flagged fraction is about 24\%.
    
\begin{figure}
        \begin{center}
            \includegraphics[width=\columnwidth]{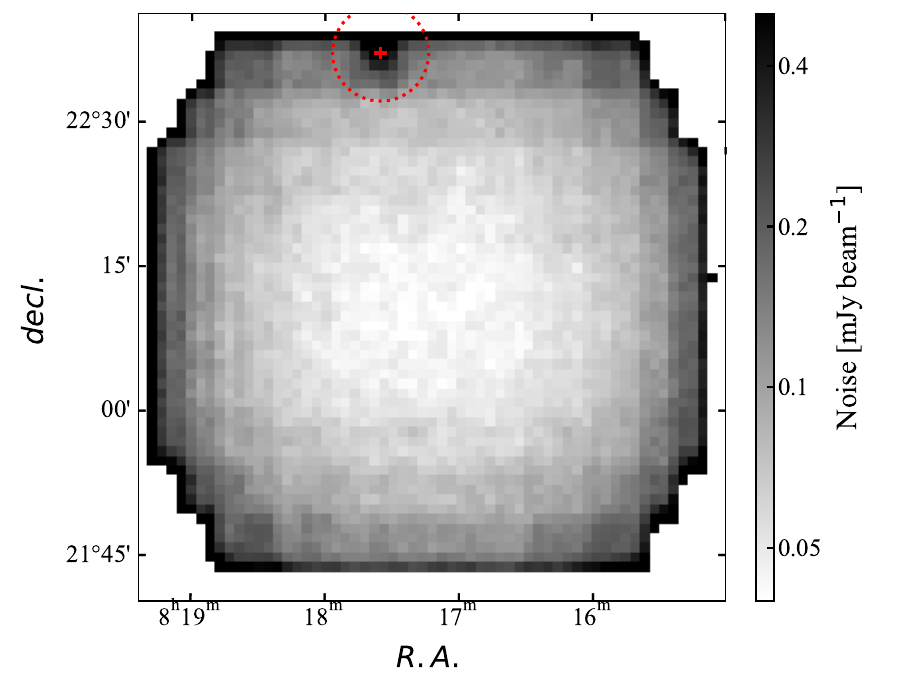}
            \caption{The spatial RMS distribution at 1.4 GHz. The red plus sign indicates the position of strongest continuum source in the field. The dotted red circle shows the extent of pixels strongly influenced by the continuum source.}\label{Fig_03}
        \end{center}
    \end{figure}

\begin{figure}
        \begin{center}
            \includegraphics[width=\columnwidth]{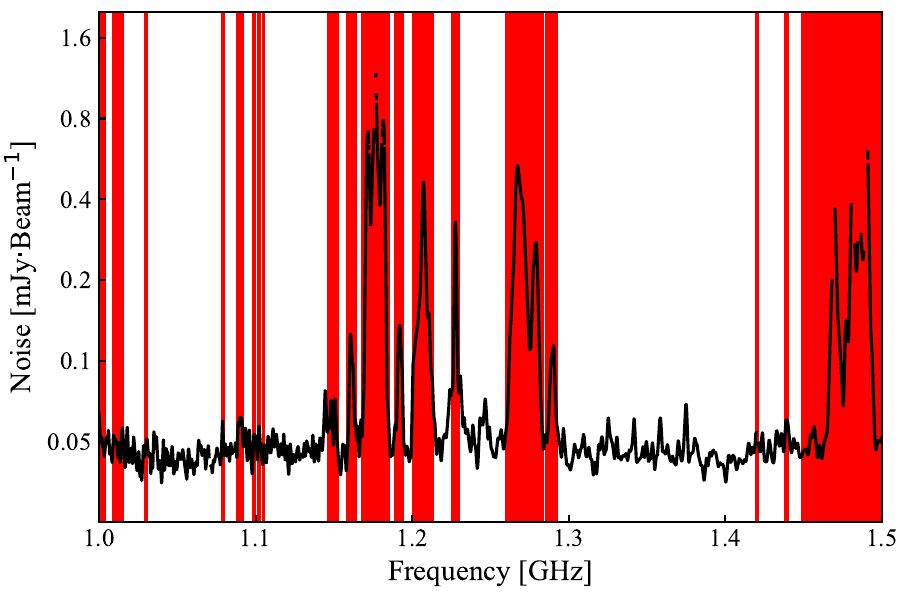}
            \caption{The RMS as a function of frequency at the center of the final cube. The red bands shows the frequency ranges strongly impacted by RFI.}\label{Fig_04}
        \end{center}
    \end{figure}

    In order to examine the noise distribution in the final data cube, we followed the procedures in \citet{2015MNRAS.452.3726H} to generate a noise cube with the same dimensions as the data cube.  For each voxel, we calculated the RMS over the channels within $\pm 150$ km~s$^{-1}$ in the same spectrum, after excluding the channels belonging to any sources detected by our source finding code (see Section \ref{Sct_04}). The noise cube was then smoothed with a Gaussian function ($\sigma=100$ km~s$^{-1}$) within $\pm 200$ km~s$^{-1}$. The resulted spatial noise distribution at 1.4 GHz is illustrated in Figure \ref{Fig_03}. The lowest noise is around the field center, as expected. There is a strong continuum source ($S_{\nu}=1.28$ Jy) at the upper edge, where the noise is strongly elevated. This impacts the detectability of nearby sources. Figure \ref{Fig_04} shows the central RMS noise as a function of frequency. As noted above, satellites are responsible for the high RMS in the frequency range 1.15 -- 1.3 GHz and at 1.48 GHz. The noise is also elevated by the aviation radar at around 1.09 GHz. Internal RFI has a modest impact on the RMS noise level, but its presence causes poor baselines and difficulty in faint source identification. Frequencies where source identification was strongly affected are marked in red in Figure \ref{Fig_04} (see also Section \ref{Sct_04}).
    
\begin{figure}
        \begin{center}
            \includegraphics[width=\columnwidth]{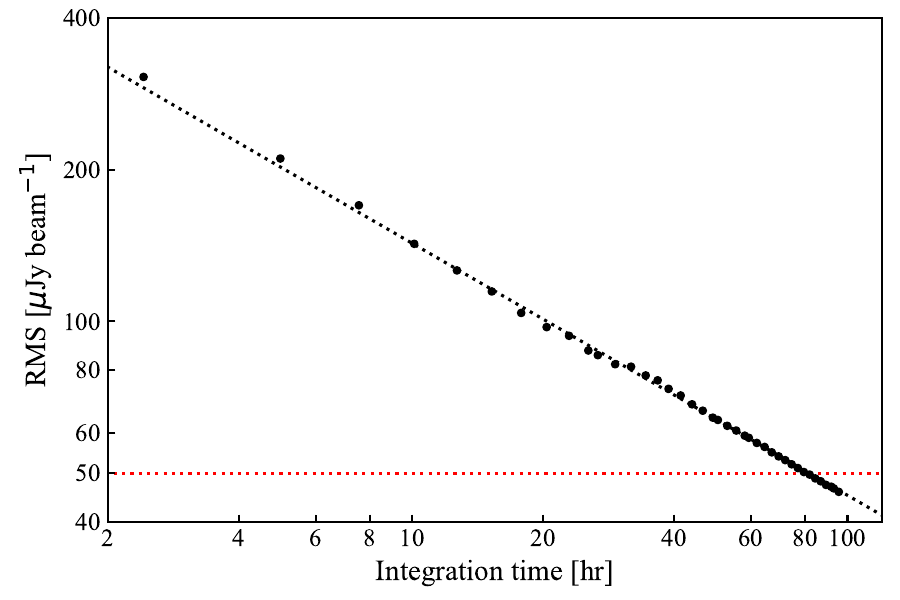}
            \caption{The noise at the center of the FUDS0 data cube as a function of integration time (black dots). The best fit line is represented by RMS $\propto T_{\rm int}^{-1/2}$ (black dotted line). The achieved target sensitivity of the FUDS survey is indicated by the red dotted line.}\label{Fig_05}
        \end{center}
    \end{figure}

    In order to measure the rate of improvement in the RMS noise as a function of integration time, we generated data cubes by adding data from consecutive observing days, each time re-generating a new noise cube. The noise cubes were smoothed by a Gaussian kernel as above. The mean RMS noise in the central pixel between frequencies of 1.39 and 1.41~GHz is shown as a function of integration time in Figure \ref{Fig_05}. The noise is proportional to $T_{\rm int}^{-1/2}$, as expected. We also achieve our target sensitivity for FUDS0, 50~$\mu$Jy, as indicated by the red dotted line in the figure.
    
\section{Source finding}\label{Sct_04}

    The noise level in our final cube is highly variable in both space and frequency (see Section \ref{Sct_03}), with non-flat spectral baselines around RFI-affected regions being problematic. The algorithm used for source finding is similar to that of our RFI finder \citep{2022PASA...39...19X}. The finder provides masks for detected signals, and flattens the spectral baselines using polynomial functions. Detections with small spatial ($<2$ pixels, equivalent to $<2$~arcmin in either $R.A.$ or $decl.$) or frequency extent ($<5$ channels, equivalent to $<38.1$~kHz in frequency) are removed. Two detection criteria were considered: 1) peak flux density $|S_{peak}| \geq 7\sigma$, where $\sigma$ is the local RMS in the smoothed spectrum; 2) $7 > |S_{peak}| \geq 5\sigma$ confirmed by additional information, including distinct double-horn feature in the line profile, with a beam-sized spatial extent, or matched redshifts from other surveys (SDSS DR15, \citealp{2019ApJS..240...23A}, AUDSOC, \citealp{2014UWAThesisH}, AUDS, \citealp{2021MNRAS.501.4550X}). Negative detections satisfying the above criteria were included for reliability analysis. The source finder did not separate galaxies joined in both spatial extent and frequency. These were later separated manually where possible.
    
    Based on the number of negative signals, we found that many false detections were around strong continuum sources or close to RFI frequencies. These detections were culled if they were within 5\,arcmin of the strong continuum source (the dotted red line in Figure \ref{Fig_03}) or within the red guard bands in Figure \ref{Fig_04}, unless they were confirmed by the additional information in the second criterion, even for the detections satisfying the first criterion.

\section{Catalogue}\label{Sct_05}

    \subsection{Extra galaxies}
    
\begin{figure}
            \begin{center}
                \includegraphics[width=\columnwidth]{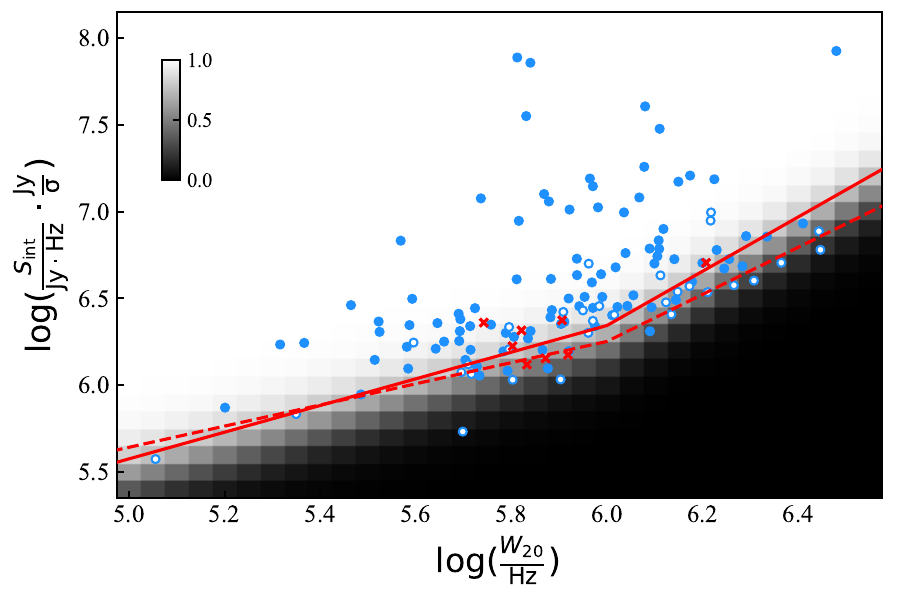}
                \caption{The distribution of FUDS0 galaxies as a function of their velocity width and noise-weighted flux. The filled and empty blue circles represent the galaxies satisfying the first and second criterion in the source finding procedure (see text), respectively. The red crosses indicate the negative detections. The gray scale image shows the two dimensional completeness function from our simulation. The solid red line is the 95\% completeness line. The $T_{\rm C} = -3$ line from modified $V/V_{\rm max}$ method is given by dashed red line.}
                \label{Fig_06}
            \end{center}
        \end{figure}

        We detect 128 galaxies\footnote{Hydroxyl megamasers (OHM) at a rest frequency of 1.66735903\,GHz have a similar spectral profile, so we cross-matched our detections with the positions and redshifted frequencies of known OHMs, or potential OHMs in the case of galaxies with known spectroscopic redshift from SDSS or DESI. We did not detect any obvious OHMs. This in line with the predictions of \citet{2021ApJ...911...38R} who estimated the number ratio between OHMs and \HI\ galaxies for the CHILES survey, which has similar sensitivity and redshift coverage, as 0.18\%.} in the FUDS0 field. Since all the FUDS0 galaxies are detected as point sources, we use a two dimensional Gaussian function with a fixed full width half maximum (FWHM) to fit the moment 0 maps to derive the sky coordinates, $R.A.$ and $decl.$ The FWHM was the beam size at corresponding frequency in the final cube, which was derived in our previous work \citep{2022PASA...39...19X}. The spatial-integrated spectrum is derived as follows:
        \begin{equation}
            S(\nu) = \frac{\Sigma_{ij} I_{ij}(\nu) w_{ij} \Delta x \Delta y}{\Sigma_{ij} w_{ij}^{2} \Delta x \Delta y}
        \end{equation}
        where $I_{ij}(\nu)$ is the intensity in pixel with indices of $i$, $j$ within the FWHM, the $w_{ij}$ is the weight from the 2D Gaussian fit, and $\Delta x \Delta y$ is the pixel solid angle . We use the Busy Function \citep{2014MNRAS.438.1176W} to fit the spatially-integrated spectrum to derive the parameters: integrated flux $S_{\rm int}$, flux density weighted frequency $\nu_{\rm cen}$, peak flux density $S_{\rm peak}$, and linewidths $W_{50}$ and $W_{20}$. The noise level for each galaxy was estimated by computing the median of the voxels in the galaxy mask in the noise cube. Considering the large variation of noise, we show the distribution of FUDS0 galaxies in linewidth versus noise normalized flux diagram (see Figure \ref{Fig_06})

        Given that some galaxies are detected close to RFI, where the spectral baseline may not be flat and the noise may be larger than elsewhere, the Busy function may underestimate the uncertainties of their parameters. Hence we employed the following method to take into account the noise and baseline around the galaxy. The method is detailed below.
        \begin{enumerate}
    
            \item The noise spectrum is extracted by subtracting the best-fit Busy function (model spectrum) from the spatially-integrated spectrum.
            
            \item The noise spectrum is shifted by $\frac{1}{30}$ of its length multiple times. An artificial spectrum is generated by adding the shifted noise spectrum to the model spectrum.
        
            \item The artificial spectrum is re-fit with a Busy function. Since a different parts of the noise spectrum will be added to the model spectrum each time, the re-fit results will reflect the impact of noise on the results.
        
            \item Steps 2 and 3 are repeated 30 times to generate a set of 30 values for each parameter. The parameter uncertainty is estimated from the standard deviation of each set.
        
        \end{enumerate}
        For flux uncertainties, we conservatively introduce a further uncertainty of 10\% to account for other sources of error, including calibration, gridding and frequency dependence. The relative error of flux parameters (including $S_{\rm int}$ and $S_{\rm peak}$ are given by the following formula:
        \begin{equation}
            \sigma_{\rm rel}' = \sqrt{\sigma_{\rm rel}^2+0.1^2} 
        \end{equation}
        where $\sigma_{\rm rel}'$ is the relative error after correction, $\sigma_{\rm rel}$ is the relative error derived by the noise shift method.

        \figsetstart
        \figsetnum{7}
        \figsettitle{Spectra for all FUDS0 galaxies.}
        
        \figsetgrpstart
        \figsetgrpnum{7.0}
        \figsetgrptitle{Examples}
        \figsetplot{Examples.pdf}
        \figsetgrpnote{Example spectra.}
        \figsetgrpend

        \figsetend

        \begin{figure*}[!h]
            \begin{center}
                \includegraphics[width=1.9\columnwidth]{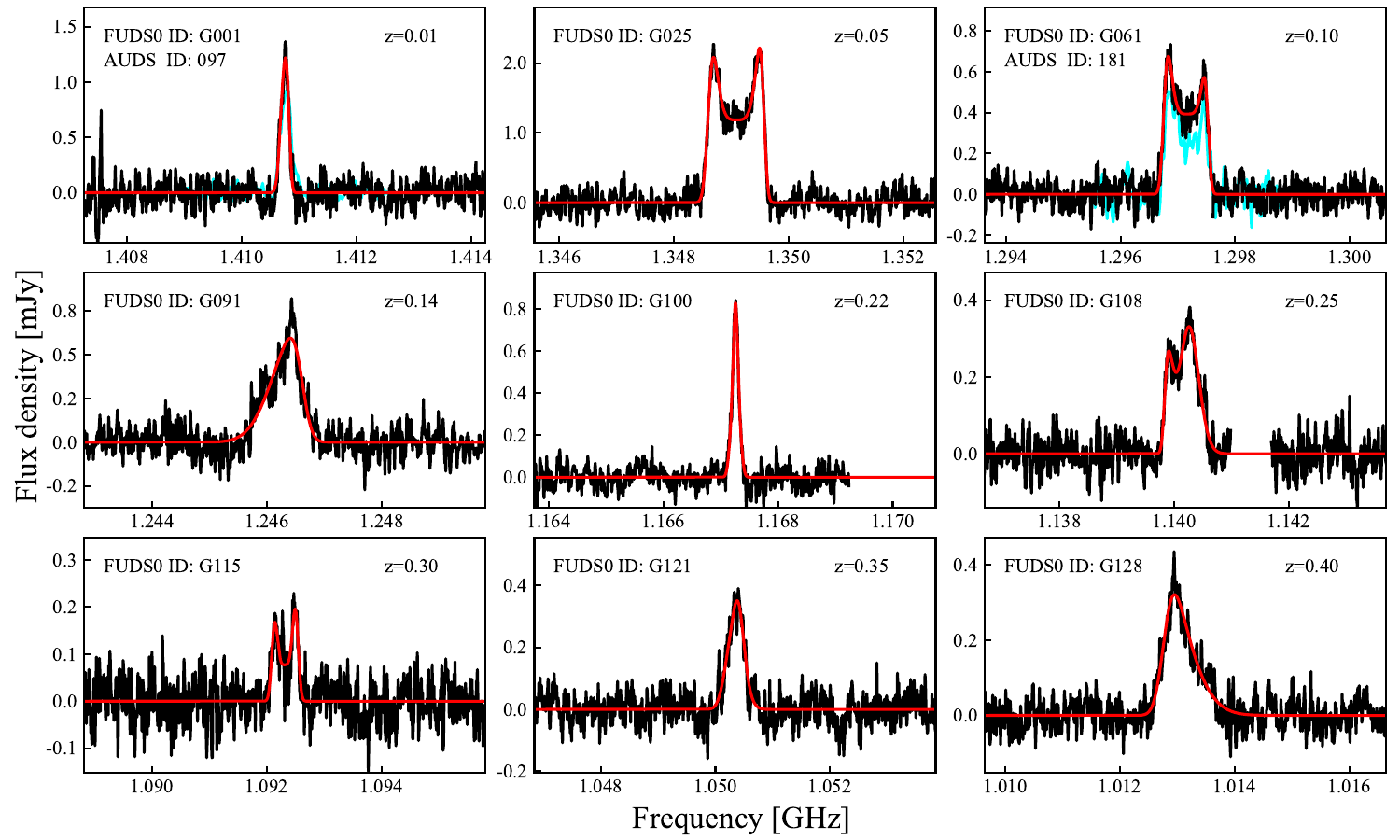}
                \caption{Example spectra at different redshifts. The red lines are the best fit Busy function. The cyan lines are the spectra from AUDS if they are available. The identifiers in FUDS0 or AUDS are given in the upper left corner, while the redshift is given in the upper right corner. Spectra for all FUDS0 galaxies can be found in the online version.}
                \label{Fig_07}
            \end{center}
        \end{figure*}

        Figure \ref{Fig_07} shows example spectra at different redshifts. The \HI\ masses for the detections are derived using \citep{2017PASA...34...52M}:
        \begin{equation}
            M_{\rm \HI} = 49.7 \cdot D_{\rm Lum}^2 \cdot S_{\rm int} ,
        \end{equation}
        where $M_{\rm \HI}$ is the \HI\ mass in $h_{70}^{-2} {\rm M_\odot}$, $D_{\rm Lum}$ is the luminosity distance of the galaxy in $h_{70}^{-1} {\rm Mpc}$, and $S_{\rm int}$ is the integrated flux in Jy$\cdot$Hz. Note that the luminosity distance was simply derived from Hubble flow based on the solar system barycentric redshift, $z_{\rm bary}$. We did not correct the distance for 1) solar system velocity relative to cosmic microwave background (CMB), 2) peculiar velocity of detected galaxies. For the galaxies partially detected at the edge of the field, we did not correct their \HI\ mass for beam shape in the cube.

        \begin{figure*}
            \begin{center}
                \includegraphics[width=1.4\columnwidth]{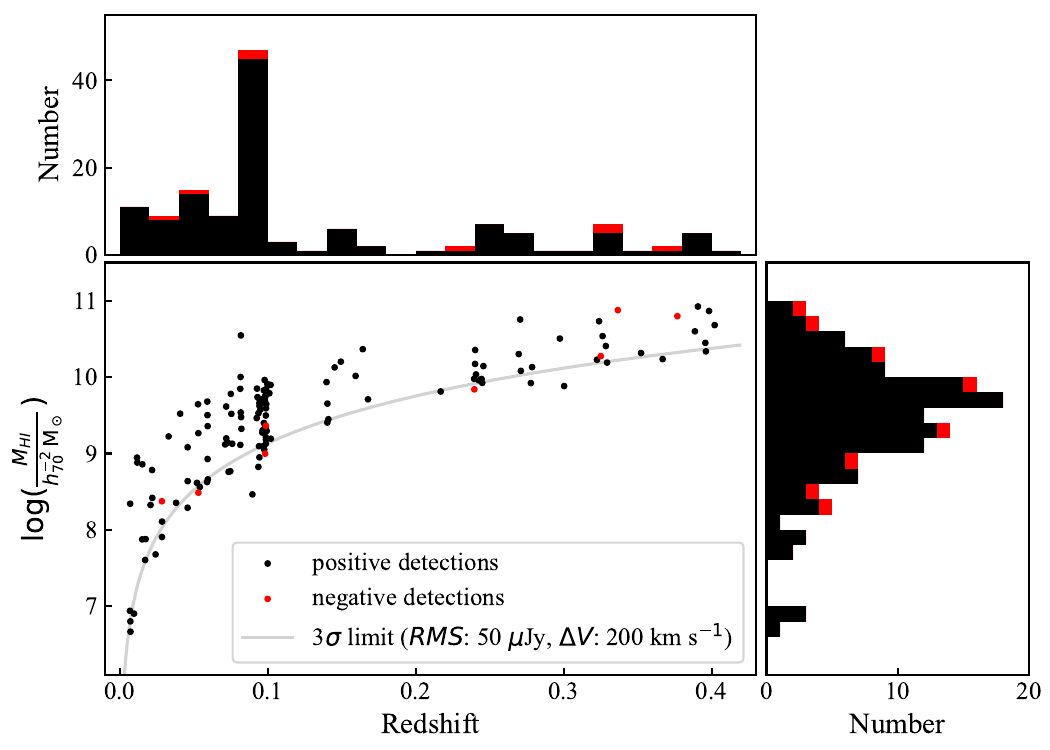}
                \caption{Redshift and HI mass distribution of detected sources in the FUDS0 field. The positive and negative detections are shown by black and red dots, respectively. The grey line represents the 3-$\sigma$ detection limit. Single-parameter histograms are also shown at the top and on the right-hand side.}
                \label{Fig_08}
            \end{center}
        \end{figure*}
        
        The HI mass distribution as a function of redshift of the 128 galaxies is shown in Figure \ref{Fig_08}. Redshifts range from 0 to 0.4, and \HI\ masses  are between $10^{6.67}$ and $10^{10.92}~h_{70}^{-2} \rm M_\odot$. Due to large-scale structure, many galaxies are detected at $z \sim 0.1$. The gap in the redshift range $0.1<z<0.25$ is due to RFI from GNSS. The grey line represents the 3-$\sigma$ integrated flux detection limit under the assumption of $RMS=50~\mu$Jy and $\Delta V = 200$~km$\cdot$s$^{-1}$. To estimate the reliability, we searched for negative detections using the same criteria as for positive detections. As a result, we found eight negative detections, shown by red dots in the figure. These negative (false) detections are evenly distributed in redshift and  mostly located close to the 3-$\sigma$ limit.
        
        The properties of the galaxies are listed in Table \ref{Tab_01}. The identifier in FUDS0 catalogue is given in column 1. Column 2 shows the J2000 right ascension ($R.A.$) and declination ($decl.$). The recession velocity ($cz_{\rm bary}$) relative to the barycenter of solar system is given in column 3. The integrated flux ($S_{\rm int}$) and peak flux density ($S_{\rm peak}$) are listed in column 4 and 5, respectively. In column 6 and 7, the linewidth $W_{20}$ and $W_{50}$ are presented. The \HI\ mass ($M_{\HI}$) is listed in column 8. The AUDS ID is given in Column 9, if the galaxy is also detected in AUDS survey.
        
        \begin{table*}
            \centering
            \caption{Properties of galaxies in the FUDS0 field. Full version is available online in mechine readable format.}
            \label{Tab_01}
            \begin{tabular}{ccccccccc}
                \hline
                ID  & $R.A.$, $decl.$ (J2000)       & $cz_{\rm bary}$ & $S_{\rm int}$ & $S_{\rm peak}$ & $W_{\rm 20}$ & $W_{\rm 50}$ & $\log(M_{\rm HI})$  & ID  \\
                FUDS0 & {\tiny HH:MM:SS.S$\pm$DD:MM:SS} & km~s$^{-1}$ & mJy~MHz       & mJy            & MHz         &     MHz &    $\log({\rm M}_\odot/h_{70}^2)$ & AUDS \\
                 (1) &                 (2) &             (3) &         (4) &          (5) &         (6) &         (7) &          (8)    &  (9)\\
                \hline
                G001 & 08:18:55.8+22:21:46 &   2052.7 (1.1) & 0.20 (0.02) &  1.22 (0.14) & 0.23 (0.02) & 0.16 (0.01) &  6.94 (0.05) & 097 \\
                G002 & 08:17:57.3+22:26:13 &   2101.0 (0.1) & 4.86 (0.49) &  9.29 (0.93) & 0.69 (0.01) & 0.54 (0.01) &  8.34 (0.04) & 098 \\
                G003 & 08:18:36.1+22:30:12 &   2148.1 (2.5) & 0.13 (0.02) &  0.63 (0.08) & 0.31 (0.02) & 0.20 (0.03) &  6.80 (0.06) & 099 \\
                G004 & 08:19:14.6+22:22:56 &   2163.7 (1.5) & 0.10 (0.02) &  1.22 (0.19) & 0.11 (0.02) & 0.07 (0.01) &  6.67 (0.06) & 100 \\
                G005 & 08:16:06.6+22:22:17 &   2877.4 (10.1) & 0.09 (0.02) &  0.18 (0.03) & 0.75 (0.10) & 0.48 (0.05) &  6.90 (0.06) &  -  \\
                G006 & 08:18:50.8+22:06:58 &   3493.8 (0.1) & 7.01 (0.70) & 14.87 (1.49) & 0.65 (0.01) & 0.57 (0.01) &  8.95 (0.04) & 106 \\
                G007 & 08:17:26.3+21:43:19 &   3543.4 (0.9) & 5.87 (0.59) &  8.10 (0.82) & 1.20 (0.01) & 0.65 (0.03) &  8.88 (0.04) &  -  \\
                G008 & 08:17:01.5+22:23:10 &   4546.3 (0.4) & 0.35 (0.04) &  1.26 (0.13) & 0.37 (0.01) & 0.28 (0.01) &  7.88 (0.05) &  -  \\
                G009 & 08:18:59.6+21:50:16 &   4578.9 (4.3) & 3.32 (0.35) &  6.54 (0.80) & 0.74 (0.05) & 0.47 (0.05) &  8.86 (0.05) & 120 \\
                G010 & 08:17:49.3+22:03:18 &   5142.2 (0.6) & 0.15 (0.02) &  0.71 (0.08) & 0.29 (0.01) & 0.20 (0.01) &  7.61 (0.04) & 121 \\
                 ... &                 ... &             ... &         ... &          ... &         ... &         ... &          ... &   ... \\
                \hline
            \end{tabular}
        \end{table*}

    \subsection{High velocity clouds}

        We also detected three HVCs near 1.42 GHz in the FUDS0 cube. They all have structure that is more extended than the beam. A simple Gaussian function was employed to model the spectrum of each HVC. The integrated intensity ($I_{\rm int}$) for each pixel was derived and used as weight to compute the coordinates, $R.A.$ and $decl.$. The column density at position $(x, y)$ was derived using the following equation \citep{2017PASA...34...52M}:
        \begin{equation}
            N_\HI(x, y) = 2.33 \times 10^{20} (1 + z)^4 I_{\rm int}(x, y) \Omega_{\rm bm} (ab)^{-1}
        \end{equation}
        where $z$ is the redshift, $I_{\rm int}(x, y)$ is the integrated intensity at position $(x, y)$ in Jy$\cdot$Hz$\cdot \Omega_{\rm bm}^{-1}$, $\Omega_{\rm bm}$ is the beam size in the cube, and $a, b$ are the beam major and minor axes in arcsec. We calculate the spatially-integrated spectrum as follows:
        \begin{equation}
            S(\nu) = 4 \ln(2)\frac{\Sigma_{ij} I_{ij}(\nu) \Delta x \Delta y}{\pi \theta^2}
        \end{equation}
        where $I_{ij}(\nu)$ is the intensity at pixel $(i,j)$, $\theta$ is the model beam size in final data cube, and $\Delta x \Delta y$ is the pixel solid angle. A Gaussian function was used to fit the spectrum to derive the HVC properties, including flux density weighted frequency ($\nu_{cen}$), integrated flux ($S_{\rm int}$) and linewidth ($W_{20}$ and $W_{50}$).
        
        The HVC properties are listed in Table \ref{Tab_02}. In column 1, the identifier is given. The integrated intensity weighted coordinates ($R.A.$ and $decl.$) in J2000 epoch are listed in column 2. Column 3 gives the velocity in the kinematic local standard of rest frame ($v_{\rm lsr}$). In column 4 and 5, the integrated flux ($S_{\rm int}$) and linewidth $W_{20}$ are listed, respectively. We present the peak column density ($N_{\rm peak}$) in column 7.

        \begin{table*}
            \centering
            \caption{Properties of HVCs in FUDS0 field.}
            \label{Tab_02}
            \begin{tabular}{cccccc}
                \hline
                 ID  & $R.A.$, $decl.$ (J2000)       & $v_{\rm lsr}$   & $S_{\rm int}$  & $W_{\rm 20}$   & $N_{\rm peak}$     \\
                 FUDS0    & {\tiny HH:MM:SS.S$\pm$DD:MM:SS} & km$\cdot$s$^{-1}$ & mJy$\cdot$MHz    & km$\cdot$s$^{-1}$     & $\times 10^{17}$\,cm$^{-2}$           \\
                     (1) &                     (2) & (3)         &            (4) &                 (5) &         (6) \\ 
                \hline
                HVC001 & 08:17:45.8+21:49:56 & 177.7 (0.5) &  0.13 (0.02)  &  44 (2) &  3.1 (1.0) \\
                HVC002 & 08:17:27.0+22:08:10 & 188.0 (0.4) &  0.14 (0.02)  &  43 (3) &  2.9 (0.3) \\
                HVC003 & 08:17:04.1+22:25:57 & 207.8 (1.2) &  0.08 (0.01)  &  32 (3) &  2.4 (0.4) \\
                \hline
            \end{tabular}
    \end{table*}
    
\section{Completeness}\label{Sct_06}

    Completeness is the fraction of galaxies detected according to specified criteria. For an \HI\ survey, completeness typically depends on a combination of parameters, including integrated flux, peak flux density, mean flux density, line width, noise level, profile shape, redshift and RFI environment. A common and accurate method to quantify completeness in a blind survey is to perform a simulation -- e.g.\ by injecting artificial sources into the data cube  \citep{2004MNRAS.350.1210Z} and measure the fraction of artificial sources recovered. However, this method is time consuming if visual inspection is required in the source finding procedure, such as in FUDS0. The other method is to derive the completeness directly from the catalogue, which is independent of the source finding procedure. \citet{2010ApJ...723.1359M} used a Gumbel function to model the distribution of galaxy linewidth at a specific \HI\ mass, estimating completeness by comparing the data with a model. \citet{2011AJ....142..170H} used the fact that the galaxy number count should be proportional to $S_{\rm int}^{-\frac{3}{2}}$ in a spatially homogeneously distributed sample with an integrated flux limit. \citet{2016AJ....151...52S} employed the modified $V/V_{\rm max}$ method \citep{2001MNRAS.324...51R} to estimate the cutoff limit for 100\% completeness. Even if valid, these methods need a larger catalogue than FUDS0 to achieve a statistical accuracy.

    \subsection{Simulation}

        Based on our source finding criteria (see Section \ref{Sct_04}), the full FUDS0 galaxy catalogue ({\it Cat}$_{\rm full}$) is divided into two sub catalogues: 1) galaxies with $SNR \geq 7$ ({\it Cat}$_{\rm high}$), which depend only on our source finding code; and 2) galaxies with $7 \geq SNR \geq 5$ ({\it Cat}$_{\rm low}$) where visual inspection is also required. In order to achieve a completeness with high accuracy for our small sample, we adopt the method as described in AUDS100 \citep{2021MNRAS.501.4550X}, in which the completeness of {\it Cat}$_{\rm high}$ is computed by simulation, then the completeness of the full catalogue is estimated by comparing {\it Cat}$_{\rm high}$ and {\it Cat}$_{\rm full}$.

        \begin{figure}
            \begin{center}
                \includegraphics[width=\columnwidth]{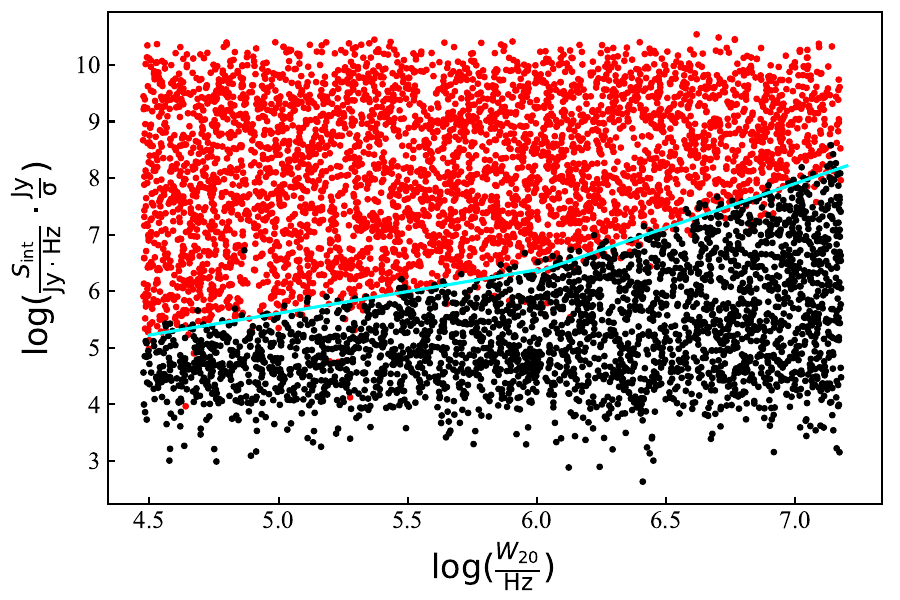}
                \caption{The distribution of artificial galaxies in the simulation. The red dots indicate the galaxies recovered by our source finding code with $SNR \ge 7$, while the black dots others. The cyan lines are the best fit by SVM to separate the detections and non-detections for the narrow ($\log(\frac{W_{20}}{\rm Hz}) < 6$) and wide ($\log(\frac{W_{20}}{\rm Hz}) \geq 6$) galaxies, respectively.}
                \label{Fig_09}
            \end{center}
        \end{figure}

\begin{figure}
            \begin{center}
                \includegraphics[width=\columnwidth]{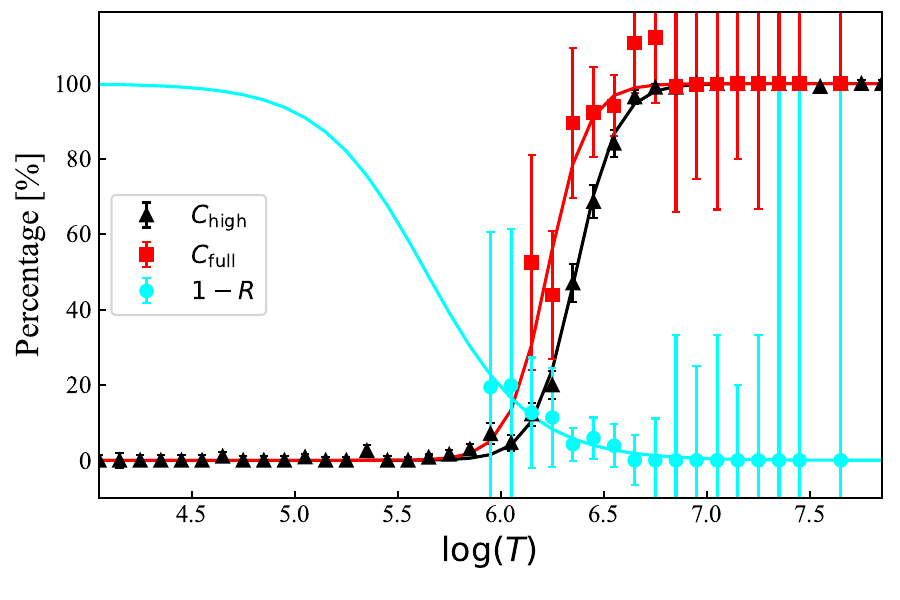}
                \caption{Completeness and false detection rate as a function of noise and linewidth normalized flux, $T$. The black line is the completeness of the artificial galaxies in our simulation, while the red line the completeness of full FUDS0 galaxies. The cyan line is the false detection rate of full FUDS0 catalogue, which is the complement of reliability to 1.}
                \label{Fig_10}
            \end{center}
        \end{figure}

        In the simulation, the Busy Function models of FUDS0 galaxies were randomly selected to generate the line profiles of the artificial galaxies, then re-scaled to randomly selected integrated fluxes and linewidths. The models were then spatially convolved with a model Gaussian beam \citep{2022PASA...39...19X} and put at randomly chosen positions in RFI and source-free regions. Our source finding code was then run on the FUDS0 cube after injecting 100 artificial galaxies. The recovered galaxies with $SNR \geq 7$ were recorded and the procedure was repeated by 100 times. As above, completeness was computed in the plane of linewidth and noise-normalized flux parameters. The simulation result is shown in Figure \ref{Fig_09}. Our code easily recovers galaxies with high noise-normalized flux. However, wider velocity width profiles with $\log(\frac{W_{20}}{\rm Hz}) \geq 6$ are much harder to detect due to the ripples in baseline. Here, we employed the method of supported vector machines (SVM) to find best lines to separate detections from non-detections of narrow ($\log(\frac{W_{20}}{\rm Hz}) < 6$) and wide ($\log(\frac{W_{20}}{\rm Hz}) \geq 6$) galaxies (the cyan lines in Figure \ref{Fig_09}), respectively. The slopes of the two lines are $\alpha = 0.77$ and $1.56$. The completeness only depends on noise and linewidth normalized flux ($T = \frac{S_{\rm int}}{\rm Jy \cdot Hz} \cdot \frac{\rm Jy}{\sigma} \cdot (\frac{\rm 10^6~Hz}{W_{20}})^\alpha$) noting again that $10^6$ Hz is the $W_{20}$ width which separates narrow and wide values for $\alpha$. 
        The detection rates were computed in bins as a function of $T$ in Figure \ref{Fig_10}. A Sigmoid function was employed to fit the completeness of {\it Cat}$_{\rm high}$.

        Considering the definition of completeness ($C = N_{\rm Det}/N_{\rm Tot}$), we have $C_{\rm full}/N_{\rm full} =  C_{\rm high}/N_{\rm high}$, which can be used to derive the completeness of $Cat_{\rm full}$ in each $T$ bin. The uncertainties were estimated by a bootstrapping technique. A Sigmoid function was used to model the completeness of our full catalogue. The best fit line is given in Figure \ref{Fig_10}. The corresponding two dimensional completeness is shown in the gray scale image in Figure \ref{Fig_06}.

        \begin{figure}
            \begin{center}
                \includegraphics[width=\columnwidth]{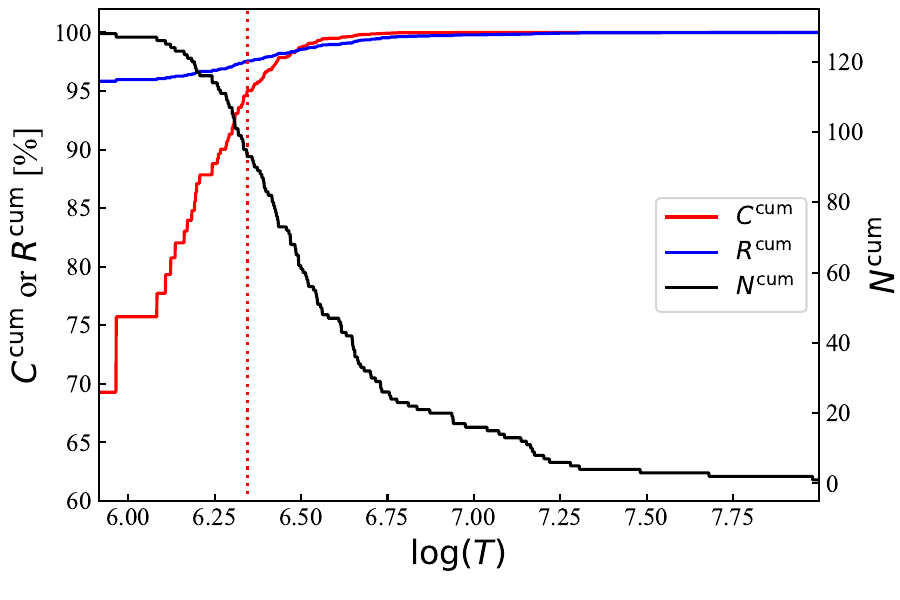}
                \caption{The cumulative completeness (red solid), cumulative reliability (blue solid) and galaxy number (black solid) as a function of noise and linewidth normalized flux, $T$. The red dotted line indicates the cutoff for the high completeness subset ($C=95\%$).}
                \label{Fig_11}
            \end{center}
        \end{figure}

        The cumulative completeness as a function of $\log(T)$ was derived with the following formula:
        \begin{equation}
            C^{\rm cum}(\log(T)) = \frac{\Sigma_i 1}{\Sigma_i \frac{1}{C_i}}
        \end{equation}
        where $T$ is noise and linewidth normalized flux, $C_i$ is the completeness of $i$th galaxy, and $i$ is the index for galaxies with $T_i \geq T$. The curve of $C^{\rm cum}(\log(T))$ is given in Figure \ref{Fig_11}. We can see that as $\log(T)$ increases, $C^{\rm cum}$ also increases, but the galaxy number decreases. Based on $\log(T) \geq 5.96$, we find that  completeness is only 69\% for full FUDS0 catalogue. Using more strict criterion, $\log(T) \geq 6.34$, we obtain a subset with high completeness of 95\%. However, the subset only consists of 93 galaxies. The limit for 95\% completeness is also given by solid red line in Figure \ref{Fig_06}. 

    \subsection{Verification}

        The modified $V/V_{\rm max}$ method of \citet{2001MNRAS.324...51R} is a data-driven method for determining completeness, and has been successfully used to characterise \HI\ catalogues in the past \citep{2004MNRAS.350.1210Z, 2016AJ....151...52S, 2021MNRAS.501.4550X}. Such an independent method helps us to verify our conclusion from the simulation. Based on the assumption of invariance in the shape of \HI\ mass function (\HIMF), the method is only dependent on the catalogue itself, and is immune to the large scale structure. The method provides an estimate for the location of the completeness limit, with the $T_{\rm C}$ statistic corresponds to probability of a given value that lies below the actual completeness limit. For example, $T_{\rm C}=-1$ for a given flux implies that the actual flux completeness limit has a 84.14\% probability of lying above that flux, and $T_{\rm C}=-3$ corresponds to 99.87\% probability (one-sided integral of normal distribution).

        The radiometer equation suggests that the completeness limit for integrated flux scales with $W^\alpha$, where $\alpha \sim 0.5$. However the baseline ripple makes it harder to detect galaxies with wide linewidths, and result in larger values for $\alpha$. The completeness limit estimated for ALFALFA from precursor observations is $\alpha=0.5$ for $W_{50}<200$~km$\cdot$s$^{-1}$ and $\alpha=1.0$ for $W_{50}\geq200$~km$\cdot$s$^{-1}$ \citep{2005AJ....130.2598G}. 
    
        An additional consideration for deep, wide redshift coverage surveys such as FUDS is the large variation of noise levels, suggesting that a local noise level should be an additional parameter for studying completeness along with integrated flux and linewidth \citep{2015MNRAS.452.3726H, 2021MNRAS.501.4550X}.
        In the FUDS noise cube (see Section \ref{Sct_03}), the local noise around each galaxy $\sigma$, is estimated from the median of local RMS values. The noise normalized flux $\log(\frac{S_{\rm int}}{\rm Jy \cdot Hz}\cdot\frac{\rm Jy}{\sigma})$, is plotted as a function of linewidth $\log(\frac{W_{\rm 20Cor}}{\rm Hz})$ in Figure \ref{Fig_06}. For the purposes of completeness estimation, we excluded an outlier that has an unusually low noise-scaled flux.  Similar to the case for ALFALFA \citep{2005AJ....130.2598G}, it seems that $\alpha$ changes slope at around $\log(\frac{W_{\rm 20Cor}}{\rm Hz}) = 6.0$. The adaptive iteratively reweighted penalized least squares (airPLS; \citealp{C4AN01061B}) method was used to estimate $\alpha$, which resulted in $\alpha=0.61$ for $\log(\frac{W_{\rm 20}}{\rm Hz}) < 6.0$ and $\alpha=1.36$ for $\log(\frac{W_{\rm 20Cor}}{\rm Hz}) \geq 6.0$. These results are also similar to those for ALFALFA, although the steeper value for $\alpha$ for wide galaxies is probably due to excess of baseline variation in FUDS0, caused by presence of RFI during the commissioning period.

\begin{figure}
            \begin{center}
                \includegraphics[width=\columnwidth]{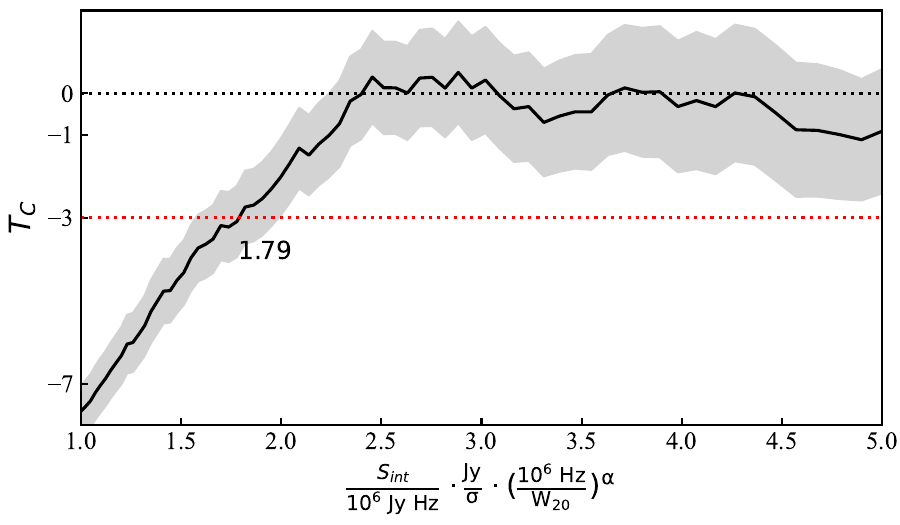}
                \caption{$T_{\rm C}$ values as a function of $Z$ for the modified $V/V_{max}$ method. The gray area displays 1-$\sigma$ uncertainty. The red dotted line indicates $T_{\rm C}=-3$.}
                \label{Fig_12}
            \end{center}
        \end{figure}

        We calculate $Z=\log(\frac{S_{\rm int}}{\rm 10^6~Jy~Hz}\cdot\frac{\rm Jy}{\sigma}\cdot(\frac{\rm 10^6~Hz}{W_{20}})^{\alpha})$ and apply the modified $V/V_{\rm max}$ method. The $T_{\rm C}$ values are calculated and shown in Figure \ref{Fig_12}. The curve shows a sharp drop at low $Z$, implying a good choice for $\alpha$. The completeness limit for $T_{\rm C}=-3$ is shown by the dashed red lines in Figure \ref{Fig_06}. The slope for wide galaxies is similar to that from the  simulation, while the slope for narrow galaxies is flatter. The discrepancy could be caused by the lack of narrow galaxies in our sample. However, the 95\% completeness line is above the completeness limit from the modified $V/V_{\rm max}$. Such a consistency confirms  our soft rolling  completeness function derived from the simulation. 

\section{Reliability}\label{Sct_07}

    Reliability is a measure of the confidence of the detections listed in a catalogue. It can be expressed by:
    \begin{equation}
        R = \frac{N_{\rm Real}}{N_{\rm Total}} = 1 - \frac{N_{\rm False}}{N_{\rm Total}}
    \end{equation}
    where $N_{\rm Real}$, $N_{\rm False}$ and $N_{\rm Total}$ are the numbers of real, false and total detections, respectively, in a catalogue. Similar to completeness, $R$ also depends on many factors, including peak flux density, flux density, linewidth and noise characteristics. For HIPASS, \citet{2004MNRAS.350.1210Z} performed a deeper re-observation for a sub-sample of fainter galaxies in the initial catalogue. $N_{\rm Real}$ was therefore available directly using the confirmed detections. For ALFALFA, \citet{2007AJ....133.2087S} performed a simulation to assess the reliability of the result based on their automated signal extractor instead of the catalogue. Artificial galaxies were inserted into a simulated data cube with white noise, and reliability was estimated from the fake detections. For AUDS, the final catalogue was formed by several steps including visual inspection \citep{2015MNRAS.452.3726H}. The number of signals only ruled out by the criterion in the last step (completeness should be greater than zeros) was used to estimate the upper limit of the reliability.

    In the FUDS0 cube, the noise is non-Gaussian, correlated across neighbouring pixels and varies in both position and frequency. We therefore attempt to identify the false detections using the data itself, rather than by generating  random noise. Firstly, four new cubes were generated by using the same gridding procedure but with different FAST data, two from independent polarization channels ($Cube_{\rm p1}$ and $Cube_{\rm p2}$), and two from timely independent and equal integration time data ($Cube_{\rm t1}$ and $Cube_{\rm t2}$). The same source finding procedures were performed on these four cubes, including: 1) running our source finding code on the cube to form a candidate list; and  2) selecting candidates satisfying the criteria mentioned in Section \ref{Sct_04}. For sources identified by the code with $S_{\rm peak} \geq 5 \sigma$ from the FUDS0 final cube, we assumed that our human judgement would give the same result as for our new cubes. Hence, if any of them were detected in the new cubes, we removed them for simplicity. Finally, we noted 7, 4, 6 and 5 new detections in these cubes, respectively. These are all false detections because they are not detected in our deeper (final) cube. Here, we assume the four new cubes have the same noise distribution in our final cube. The number and properties of these false detections can be used to estimate the reliability of FUDS0 catalogue, if we use $T$ (also used in completeness study in Section \ref{Sct_06}) to remove the effects of different noise level between the new cubes and the final cube.

    The same procedure was used for extracting the properties of 22 false detections from our 128 galaxies.  The false detection rates were computed as a function of $T$ by dividing the average number of the false detections in the new cubes by the number of FUDS0 galaxies in each bin (see Figure \ref{Fig_10}). The uncertainties were estimated by the bootstrapping method. It is reasonable to find a higher false detection rate in lower $T$ region. A Sigmoid function was also used to fit the data points. Then, the reliability can be derived by subtracting the false detection rate from unity.

    We derived the cumulative reliability as a function of $\log(T)$ using the following formula:
    \begin{equation}
        R^{\rm cum}(\log(T)) = \frac{\Sigma_i R_i}{\Sigma_i 1}
    \end{equation}
    where $T$ is noise and linewidth normalized flux, $R_i$ is the reliability of $i$th galaxy, and $i$ is the index for galaxies with $T_i \geq T$. The $R^{\rm cum}(\log(T))$ is also given in Figure \ref{Fig_11}. With increasing $\log(T)$, the cumulative reliability shows the same trend as cumulative completeness. Based on same criterion, $\log(T) \geq 5.96$, we found a high reliability of 96\% for our full FUDS0 catalogue. For the high completeness subset ($\log(T) \geq 6.34$), the reliability is even higher, 98\%.

    In the source finding procedure, we recorded negative sources, which can  also be used to estimate the false detection rate. In this case, the reliability for the full FUDS0 catalogue is $R = 1 - 8/128 = 94\%$, which is close to 96\% reliability for the full catalogue estimated above. The consistency confirms that the method for studying reliability above is reasonable.

\section{Comparison with previous works}\label{Sct_08}

    \subsection{NVSS}

        \begin{figure}
            \begin{center}
                \includegraphics[width=0.8\columnwidth]{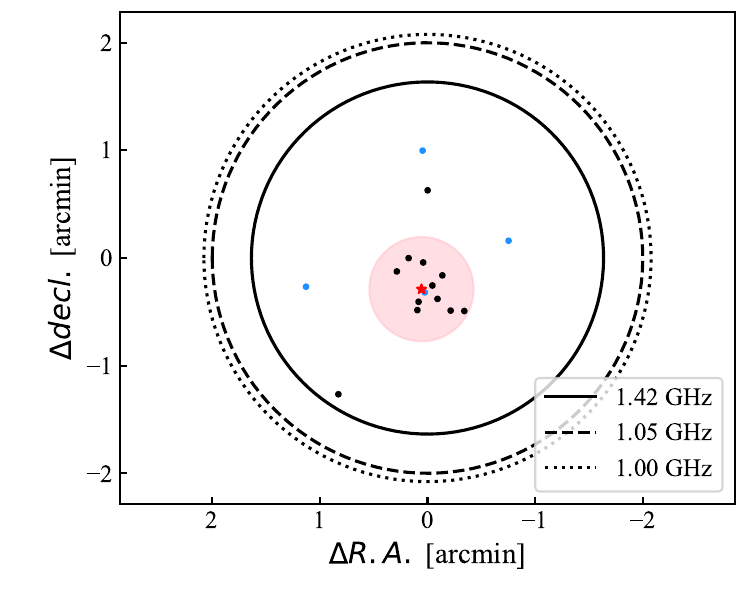}
                \includegraphics[width=0.8\columnwidth]{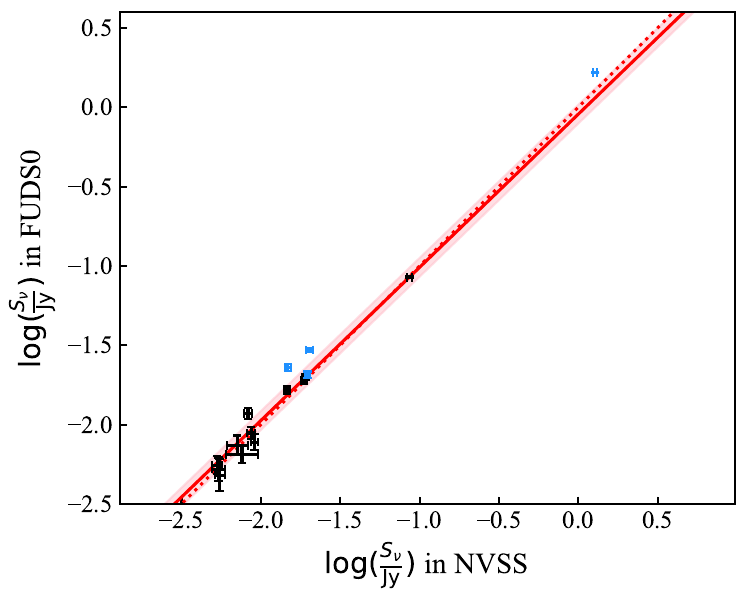}
                \caption{Upper panel: the position offset of FUDS0 from NVSS. The black dots are sources in the high-sensitivity region in the FUDS0 field, while the blue dots are sources in the low sensitivity region. The red star indicates the mean offset of black dots. The pink area is 1-$\sigma$ uncertainty of position. The beam sizes in FUDS0 final cube are given at different frequencies (1.42, 1.05, 1.00\,GHz). Lower panel: comparison of flux density between FUDS0 and NVSS. The color of the data points have the same meaning in upper panel. The regression line and 1-$\sigma$ uncertainty are given by red line and pink area, respectively. The red dotted line indicates the 1:1 line.}
                \label{Fig_13}
            \end{center}
        \end{figure}

        The NVSS survey \citep{1998AJ....115.1693C} targets continuum sources in the northern hemisphere at 1.4\,GHz. The D and DnC configurations of the VLA array was employed to uniformly map the sky southern to $decl. = -40$\,deg, at a spatial resolution of 45\,arcsec. The survey produced a catalogue of $\sim 2 \times 10^6$ discrete sources stronger than $S \approx 2.5$\,mJy. Comparison between NVSS and FUDS0 allows an insight on the accuracy of position and flux density in our catalogue. We generated a new cube with continuum emission. After removing known galaxies, the flux density was computed by averaging the RFI-free spectra between 1.31 and 1.41\,GHz. There are 57 NVSS source in FUDS0 field. However, blending is serious due to the large beam size of FAST. Based on the criterion of angular separation of 3\,arcmin, we found 16 isolated sources, for which 12 are in the high-sensitivity region ($RMS < 3 \times 50~\mu$Jy~beam$^{-1}$) and 4 in the low sensitivity region ($RMS \geq 3 \times 50~\mu$Jy~beam$^{-1}$). Two dimensional Gaussian fits to the flux density map were used to derive positions and flux densities.

        The upper panel of Figure \ref{Fig_13} shows that all the position offsets are within the FAST beam size. The sources in the high-sensitivity region have a smaller offset than those in low sensitivity region. Because the low sensitivity region is located at the edge of the FUDS0 field, where a larger noise variation would result in greater uncertainties for best fit parameters. Based on the sources in the high-sensitivity region, we derived a mean offset of 0.30\,arcmin ($\Delta R.A. = 0.06$\,arcmin, $\Delta decl. = -0.29$\,arcmin) and a 1-$\sigma$ uncertainty of 0.48\,arcmin.
        
        A comparison of flux density between NVSS and FUDS0 is shown in the lower panel of Figure \ref{Fig_13}. Considering the impact on flux density from highly variable noise at field edge, we only performed a linear regression for the sources in the high-sensitivity region. The best fit line and 1-$\sigma$ uncertainty are given by red line and the pink area in the figure, respectively. The separation between best fit and identity line is within 1-$\sigma$ uncertainty. The consistency indicates a high accuracy of our calibration method.

    \subsection{AUDS}\label{Sct_08_02}
    
        The AUDS survey is a deep (noise level of 75\,$\mu$Jy$\cdot$beam$^{-1}$), blind survey for extragalactic \HI\ at redshifts $z<0.16$  with the Arecibo telescope. AUDS \citep{2021MNRAS.501.4550X} detected 152 galaxies in \HI\ in the GAL2577 field, which partially overlaps with FUDS0 (see Figure \ref{Fig_01}). 61 out of these 152 galaxies are in the overlapping region, 33 of which are also detected in FUDS0. 20 galaxies are located at the high-sensitivity overlapping region ($RMS<3 \times 75~\mu$Jy~beam$^{-1}$ for AUDS galaxies and $RMS<3 \times 50~\mu$Jy~beam$^{-1}$ for FUDS galaxies).
        
\begin{figure}
            \begin{center}
                \includegraphics[width=0.8\columnwidth]{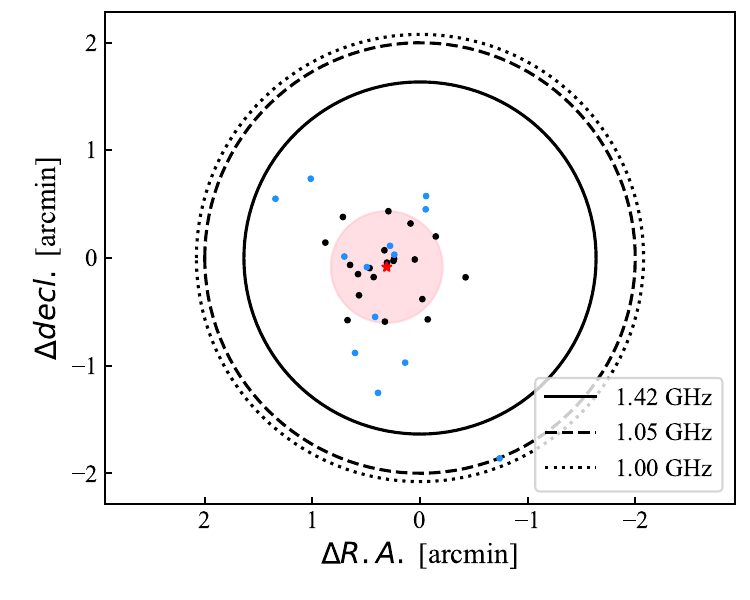}
                \includegraphics[width=0.8\columnwidth]{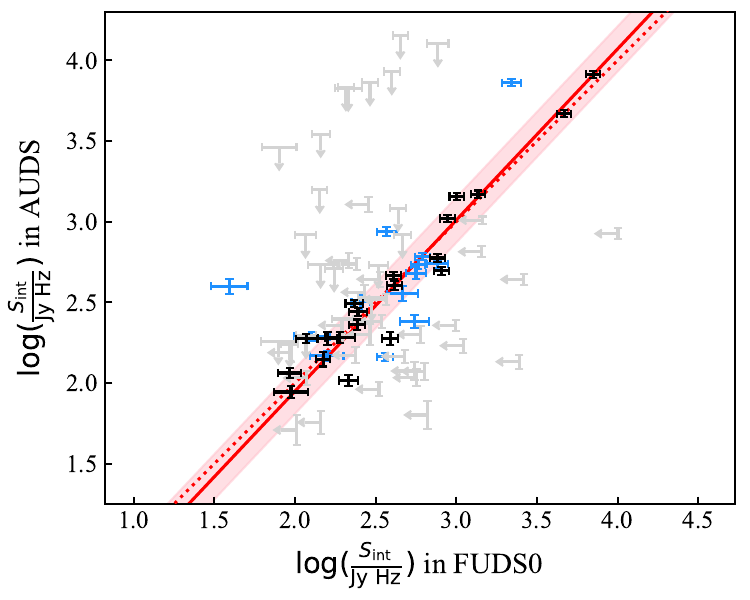}
                \caption{Upper panel: position offsets of FUDS0 galaxies from their counterparts in AUDS. The black dots represent the galaxies in the high-sensitivity overlapping region, while blue denotes the remaining galaxies. The mean offset and position uncertainty are given by red star and pink area, respectively. The beam sizes in FUDS0 cube are given at 1.4, 1.05 and 1.00\,GHz. Lower panel: comparison of integrated flux between AUDS and FUDS0. The \HI\ galaxies detected in both surveys are shown with plus signs. The black pluses are those detected in the high-sensitivity overlapping region, while the blue pluses denote the remainder. The linear regression line for black pluses is given by the solid red line, while the 1-$\sigma$ uncertainty is shown by the pink area. The dotted red line represents the identity line. The grey arrows are the upper limits of integrated flux for the galaxies detected in one survey but missed in the other.
                }
                \label{Fig_14}
            \end{center}
        \end{figure}

        The position offsets of FUDS0 galaxies from AUDS are displayed in the upper panel of Figure \ref{Fig_14}. The mean offset was derived as 0.32\,arcmin ($\Delta R.A. = 0.31$\,arcmin, $\Delta decl. = 0.08$\,arcmin) using the galaxies in the high-sensitivity region. The 1-$\sigma$ uncertainty of position is estimated to be 0.52\,arcmin. The mean offset angular distance and position uncertainty are consistent with that from the comparison with NVSS.
        
        The integrated fluxes of these galaxies are compared in the lower panel of Figure \ref{Fig_14}. For FUDS0, as noted above, we take into account an extra 10\% error due to uncertainties in calibration and imaging. This is ignored in AUDS100, which gives rise to a smaller apparent error bars. A linear regression was performed for the integrated fluxes of the galaxies in the high-sensitivity region. The best fitting line is illustrated by the red solid line, and the identity line is represented by the red dotted line. We find no systematic measurement difference between the two surveys. For the two outlier measurements, one has a low AUDS signal-to-noise ratio and the other one has a FUDS measurement affected by RFI.

        In the overlapping area, we have 28 AUDS galaxies that are not detected in FUDS. We used a top-hat line profile with the AUDS values for linewidth and a peak flux density given by three times the local noise in the FUDS0 cube to calculate the upper limits to the FUDS0 integrated fluxes. These limits are also shown in Figure \ref{Fig_14}. Most of them are below the FUDS0 detection limit. Of the 12 galaxies above or near (1-$\sigma$ consistent with) the regression line, 5 galaxies are located in the GNSS band (1.2 - 1.3~GHz), and 7 galaxies are near the spikes in the flagged fraction between 1.3 and 1.42~GHz, which is due to the compressor RFI. In the FUDS0 catalogue, we have 20 galaxies that are not detected in AUDS. The upper limits of their integrated flux for these galaxies are estimated by the same method as described above and shown in Figure \ref{Fig_14}. All of these 20 galaxies are below the detection limit in AUDS.

        Considering the larger sky area, the wider redshift coverage and higher sensitivity of the FUDS0 observations, we predicted around 152 detections based on \HI\ detections in the AUDS GAL2577 field), but only found 128. This is mainly due to the reduced sensitivity caused by the internal RFI that was generated from the compressor during the commissioning observation period. This internal RFI affected the whole bandwidth (500~MHz, detailed in \citealp{2022PASA...39...19X}). Although this also caused spectral baseline problems in some specific frequencies, our RMS and flux measurements, respectively, are consistent with what we expected (see Section \ref{Sct_03} and \ref{Sct_03}). Since the internal RFI was mitigated in 2021 July, we expect higher detection rates in the full FUDS project. Figure \ref{Fig_14} shows that 33 galaxies were detected in both AUDS and FUDS, and 12 galaxies were missed by FUDS due to bad baselines. We estimate that we should have detected around 175 ($128 \times (33 + 12) / 33$) \HI\ galaxies in the FUDS0 field if the internal RFI did not exist in our data. Based on these statistics for the FUDS0 field, the full FUDS catalogue over six fields is predicted to consist of $\sim$1000 \HI\ detections.

\section{Summary}\label{Sct_09}

    We carried out a  FAST Ultra-Deep Survey (FUDS) to explore the evolution of \HI\ content in galaxies across cosmic time. Observations for the FUDS0 field, the first of six, has been completed. The results are summarised below. 
    
    \begin{enumerate}
    
        \item The FUDS0 field ($\sim 0.72$ deg$^2$) was mapped using the FAST radio telescope for an on-source integration time of 95 hours. The survey covers a redshift range from 0.05 to 0.42 for the \HI\ emission line. High sensitivity and low noise ($\sim 50~\mu$Jy~beam$^{-1}$) was achieved at the centre of the field, rising to $\sim 300$~mJy~beam$^{-1}$ toward the edge of the field due to less integration time. The channel spacing is 7.63 kHz, and frequency resolution is 22.9 kHz after Hanning smoothing.
        
        \item We detected 128 \HI\ galaxies in the redshift range from 0 to 0.4. The \HI\ mass range for the detected galaxies is $6.67 \leq \log(\frac{M_{\HI}}{h_{70}^{-2} M_{\odot}}) \leq 10.92$. Six out of 128 \HI\ galaxies are found at $z>0.38$ \citep{2024ApJ...966L..36X}, where which no $\HI$ galaxies have previously been directly detected, other than a possible detection via strong lensing. We also detected three high velocity clouds ($v_{lsr} \geq 177$\,km$\cdot$s$^{-1}$) in the vicinity of the Milky Way. Their peak column densities are $N_{\rm HI} \leq 3.1 \times 10^{17}$~cm$^{-2}$.

        \item The completeness was derived as a function of noise and linewidth normalized flux, $T = \frac{S_{\rm int}}{\rm Jy \cdot Hz} \cdot \frac{\rm Jy}{\sigma} \cdot (\frac{\rm 10^6~Hz}{W_{20}})^\alpha$, based on the recovery of artificial galaxies. The completeness of the full FUDS0 catalogue is computed to be 69\% with a criterion of $\log(T) > 5.96$.

        \item Reliability was also derived as a function of $T$ based on the distribution of false detections. The reliability of the full FUDS0 catalogue is found to be 96\% with same criterion for completeness, $\log(T) > 5.96$.
        
        \item Comparisons of positions and integrated fluxes (or flux densities) were performed between FUDS0 and previous work (including NVSS and AUDS). The consistency demonstrates that the accuracy of our calibration method and data reduction pipeline for the FUDS project.
    
    \end{enumerate}

\begin{acknowledgments}
    This work made use of the data from FAST (Five-hundred-meter Aperture Spherical radio Telescope). FAST is a Chinese national mega-science facility, operated by the National Astronomical Observatories, Chinese Academy of Sciences. The work is supported by the National Key R\&D Program of China under grant number 2018YFA0404703, the Science and Technology Innovation Program of Hunan Province under grant number 2024JC0001, and the FAST Collaboration. Parts of this research were supported by the Australian Research Council Centre of Excellence for All Sky Astrophysics in 3 Dimensions (ASTRO 3D), through project number CE170100013.
\end{acknowledgments}

%

\vspace{5mm}
\facilities{FAST}


\software{astropy \citep{2013A&A...558A..33A, 2018AJ....156..123A}}





\bibliography{References}{}

\begin{thebibliography}{}
\expandafter\ifx\csname natexlab\endcsname\relax\def\natexlab#1{#1}\fi
\providecommand{\url}[1]{\href{#1}{#1}}
\providecommand{\dodoi}[1]{doi:~\href{http://doi.org/#1}{\nolinkurl{#1}}}
\providecommand{\doeprint}[1]{\href{http://ascl.net/#1}{\nolinkurl{http://ascl.net/#1}}}
\providecommand{\doarXiv}[1]{\href{https://arxiv.org/abs/#1}{\nolinkurl{https://arxiv.org/abs/#1}}}

\bibitem[{{Aguado} {et~al.}(2019){Aguado}, {Ahumada}, {Almeida}, {Anderson}, {Andrews}, {Anguiano}, {Aquino Ort{\'\i}z}, {Arag{\'o}n-Salamanca}, {Argudo-Fern{\'a}ndez}, {Aubert}, {Avila-Reese}, {Badenes}, {Barboza Rembold}, {Barger}, {Barrera-Ballesteros}, {Bates}, {Bautista}, {Beaton}, {Beers}, {Belfiore}, {Bernardi}, {Bershady}, {Beutler}, {Bird}, {Bizyaev}, {Blanc}, {Blanton}, {Blomqvist}, {Bolton}, {Boquien}, {Borissova}, {Bovy}, {Brandt}, {Brinkmann}, {Brownstein}, {Bundy}, {Burgasser}, {Byler}, {Cano Diaz}, {Cappellari}, {Carrera}, {Cervantes Sodi}, {Chen}, {Cherinka}, {Choi}, {Chung}, {Coffey}, {Comerford}, {Comparat}, {Covey}, {da Silva Ilha}, {da Costa}, {Dai}, {Damke}, {Darling}, {Davies}, {Dawson}, {de Sainte Agathe}, {Deconto Machado}, {Del Moro}, {De Lee}, {Diamond-Stanic}, {Dom{\'\i}nguez S{\'a}nchez}, {Donor}, {Drory}, {du Mas des Bourboux}, {Duckworth}, {Dwelly}, {Ebelke}, {Emsellem}, {Escoffier}, {Fern{\'a}ndez-Trincado}, {Feuillet}, {Fischer}, {Fleming}, {Fraser-McKelvie}, {Freischlad},
  {Frinchaboy}, {Fu}, {Galbany}, {Garcia-Dias}, {Garc{\'\i}a-Hern{\'a}ndez}, {Garma Oehmichen}, {Geimba Maia}, {Gil-Mar{\'\i}n}, {Grabowski}, {Gu}, {Guo}, {Ha}, {Harrington}, {Hasselquist}, {Hayes}, {Hearty}, {Hernandez Toledo}, {Hicks}, {Hogg}, {Holley-Bockelmann}, {Holtzman}, {Hsieh}, {Hunt}, {Hwang}, {Ibarra-Medel}, {Jimenez Angel}, {Johnson}, {Jones}, {J{\"o}nsson}, {Kinemuchi}, {Kollmeier}, {Krawczyk}, {Kreckel}, {Kruk}, {Lacerna}, {Lan}, {Lane}, {Law}, {Lee}, {Li}, {Lian}, {Lin}, {Lin}, {Lintott}, {Long}, {Longa-Pe{\~n}a}, {Mackereth}, {de la Macorra}, {Majewski}, {Malanushenko}, {Manchado}, {Maraston}, {Mariappan}, {Marinelli}, {Marques-Chaves}, {Masseron}, {Masters}, {McDermid}, {Medina Pe{\~n}a}, {Meneses-Goytia}, {Merloni}, {Merrifield}, {Meszaros}, {Minniti}, {Minsley}, {Muna}, {Myers}, {Nair}, {Correa do Nascimento}, {Newman}, {Nitschelm}, {Olmstead}, {Oravetz}, {Oravetz}, {Ortega Minakata}, {Pace}, {Padilla}, {Palicio}, {Pan}, {Pan}, {Parikh}, {Parker}, {Peirani}, {Penny}, {Percival},
  {Perez-Fournon}, {Peterken}, {Pinsonneault}, {Prakash}, {Raddick}, {Raichoor}, {Riffel}, {Riffel}, {Rix}, {Robin}, {Roman-Lopes}, {Rose}, {Ross}, {Rossi}, {Rowlands}, {Rubin}, {S{\'a}nchez}, {S{\'a}nchez-Gallego}, {Sayres}, {Schaefer}, {Schiavon}, {Schimoia}, {Schlafly}, {Schlegel}, {Schneider}, {Schultheis}, {Seo}, {Shamsi}, {Shao}, {Shen}, {Shetty}, {Simonian}, {Smethurst}, {Sobeck}, {Souter}, {Spindler}, {Stark}, {Stassun}, {Steinmetz}, {Storchi-Bergmann}, {Stringfellow}, {Su{\'a}rez}, {Sun}, {Taghizadeh-Popp}, {Talbot}, {Tayar}, {Thakar}, {Thomas}, {Tissera}, {Tojeiro}, {Troup}, {Unda-Sanzana}, {Valenzuela}, {Vargas-Maga{\~n}a}, {V{\'a}zquez-Mata}, {Wake}, {Weaver}, {Weijmans}, {Westfall}, {Wild}, {Wilson}, {Woods}, {Yan}, {Yang}, {Zamora}, {Zasowski}, {Zhang}, {Zheng}, {Zheng}, {Zhu}, {Zinn}, \& {Zou}}]{2019ApJS..240...23A}
{Aguado}, D.~S., {Ahumada}, R., {Almeida}, A., {et~al.} 2019, \apjs, 240, 23, \dodoi{10.3847/1538-4365/aaf651}

\bibitem[{{Astropy Collaboration} {et~al.}(2013){Astropy Collaboration}, {Robitaille}, {Tollerud}, {Greenfield}, {Droettboom}, {Bray}, {Aldcroft}, {Davis}, {Ginsburg}, {Price-Whelan}, {Kerzendorf}, {Conley}, {Crighton}, {Barbary}, {Muna}, {Ferguson}, {Grollier}, {Parikh}, {Nair}, {Unther}, {Deil}, {Woillez}, {Conseil}, {Kramer}, {Turner}, {Singer}, {Fox}, {Weaver}, {Zabalza}, {Edwards}, {Azalee Bostroem}, {Burke}, {Casey}, {Crawford}, {Dencheva}, {Ely}, {Jenness}, {Labrie}, {Lim}, {Pierfederici}, {Pontzen}, {Ptak}, {Refsdal}, {Servillat}, \& {Streicher}}]{2013A&A...558A..33A}
{Astropy Collaboration}, {Robitaille}, T.~P., {Tollerud}, E.~J., {et~al.} 2013, \aap, 558, A33, \dodoi{10.1051/0004-6361/201322068}

\bibitem[{{Astropy Collaboration} {et~al.}(2018){Astropy Collaboration}, {Price-Whelan}, {Sip{\H{o}}cz}, {G{\"u}nther}, {Lim}, {Crawford}, {Conseil}, {Shupe}, {Craig}, {Dencheva}, {Ginsburg}, {VanderPlas}, {Bradley}, {P{\'e}rez-Su{\'a}rez}, {de Val-Borro}, {Aldcroft}, {Cruz}, {Robitaille}, {Tollerud}, {Ardelean}, {Babej}, {Bach}, {Bachetti}, {Bakanov}, {Bamford}, {Barentsen}, {Barmby}, {Baumbach}, {Berry}, {Biscani}, {Boquien}, {Bostroem}, {Bouma}, {Brammer}, {Bray}, {Breytenbach}, {Buddelmeijer}, {Burke}, {Calderone}, {Cano Rodr{\'\i}guez}, {Cara}, {Cardoso}, {Cheedella}, {Copin}, {Corrales}, {Crichton}, {D'Avella}, {Deil}, {Depagne}, {Dietrich}, {Donath}, {Droettboom}, {Earl}, {Erben}, {Fabbro}, {Ferreira}, {Finethy}, {Fox}, {Garrison}, {Gibbons}, {Goldstein}, {Gommers}, {Greco}, {Greenfield}, {Groener}, {Grollier}, {Hagen}, {Hirst}, {Homeier}, {Horton}, {Hosseinzadeh}, {Hu}, {Hunkeler}, {Ivezi{\'c}}, {Jain}, {Jenness}, {Kanarek}, {Kendrew}, {Kern}, {Kerzendorf}, {Khvalko}, {King}, {Kirkby}, {Kulkarni},
  {Kumar}, {Lee}, {Lenz}, {Littlefair}, {Ma}, {Macleod}, {Mastropietro}, {McCully}, {Montagnac}, {Morris}, {Mueller}, {Mumford}, {Muna}, {Murphy}, {Nelson}, {Nguyen}, {Ninan}, {N{\"o}the}, {Ogaz}, {Oh}, {Parejko}, {Parley}, {Pascual}, {Patil}, {Patil}, {Plunkett}, {Prochaska}, {Rastogi}, {Reddy Janga}, {Sabater}, {Sakurikar}, {Seifert}, {Sherbert}, {Sherwood-Taylor}, {Shih}, {Sick}, {Silbiger}, {Singanamalla}, {Singer}, {Sladen}, {Sooley}, {Sornarajah}, {Streicher}, {Teuben}, {Thomas}, {Tremblay}, {Turner}, {Terr{\'o}n}, {van Kerkwijk}, {de la Vega}, {Watkins}, {Weaver}, {Whitmore}, {Woillez}, {Zabalza}, \& {Astropy Contributors}}]{2018AJ....156..123A}
{Astropy Collaboration}, {Price-Whelan}, A.~M., {Sip{\H{o}}cz}, B.~M., {et~al.} 2018, \aj, 156, 123, \dodoi{10.3847/1538-3881/aabc4f}

\bibitem[{Baek {et~al.}(2015)Baek, Park, Ahn, \& Choo}]{C4AN01061B}
Baek, S.-J., Park, A., Ahn, Y.-J., \& Choo, J. 2015, Analyst, 140, 250, \dodoi{10.1039/C4AN01061B}

\bibitem[{{Baker} {et~al.}(2018){Baker}, {Blyth}, {Holwerda}, \& {LADUMA Team}}]{2018AAS...23123107B}
{Baker}, A.~J., {Blyth}, S., {Holwerda}, B.~W., \& {LADUMA Team}. 2018, in American Astronomical Society Meeting Abstracts, Vol. 231, American Astronomical Society Meeting Abstracts \#231, 231.07

\bibitem[{{Barnes} {et~al.}(2001){Barnes}, {Staveley-Smith}, {de Blok}, {Oosterloo}, {Stewart}, {Wright}, {Banks}, {Bhathal}, {Boyce}, {Calabretta}, {Disney}, {Drinkwater}, {Ekers}, {Freeman}, {Gibson}, {Green}, {Haynes}, {te Lintel Hekkert}, {Henning}, {Jerjen}, {Juraszek}, {Kesteven}, {Kilborn}, {Knezek}, {Koribalski}, {Kraan-Korteweg}, {Malin}, {Marquarding}, {Minchin}, {Mould}, {Price}, {Putman}, {Ryder}, {Sadler}, {Schr{\"o}der}, {Stootman}, {Webster}, {Wilson}, \& {Ye}}]{2001MNRAS.322..486B}
{Barnes}, D.~G., {Staveley-Smith}, L., {de Blok}, W.~J.~G., {et~al.} 2001, \mnras, 322, 486, \dodoi{10.1046/j.1365-8711.2001.04102.x}

\bibitem[{{Blyth} {et~al.}(2016){Blyth}, {Baker}, {Holwerda}, {Bouchard}, {Catinella}, {Chemin}, {Cunnama}, {Dav{\'e}}, {Faltenbacher}, {February}, {Fern{\'a}ndez}, {Gawiser}, {Heywood}, {Kere{\v{s}}}, {Kl{\"o}ckner}, {Lah}, {Lochner}, {Maddox}, {Makhathini}, {Moodley}, {Morganti}, {Obreschkow}, {Oh}, {Pisano}, {Popping}, {Popping}, {Ravindranath}, {Schinnerer}, {Sheth}, {Skelton}, {Smith}, {Srianand}, {Staveley-Smith}, {Vaccari}, {Vaisanen}, {Walter}, {Rawlings}, {Bassett}, {Bershady}, {Briggs}, {Crawford}, {Cress}, {Darling}, {Deane}, {de Blok}, {Elson}, {Frank}, {Henning}, {Hess}, {Hughes}, {Jarvis}, {Kannappan}, {Katz}, {Kraan-Korteweg}, {Lehnert}, {Leroy}, {Meurer}, {Meyer}, {Pisano}, {Schr{\"o}der}, {Smirnov}, {Somerville}, {Stewart}, {van der Heyden}, {Verheijen}, {Wilcots}, {Williams}, {Woudt}, {Wu}, {Zwaan}, {Zwart}, {Oosterloo}, \& {van Drie}}]{2016mks..confE...4B}
{Blyth}, S., {Baker}, A.~J., {Holwerda}, B., {et~al.} 2016, in MeerKAT Science: On the Pathway to the SKA, 4

\bibitem[{{Condon} {et~al.}(1998){Condon}, {Cotton}, {Greisen}, {Yin}, {Perley}, {Taylor}, \& {Broderick}}]{1998AJ....115.1693C}
{Condon}, J.~J., {Cotton}, W.~D., {Greisen}, E.~W., {et~al.} 1998, \aj, 115, 1693, \dodoi{10.1086/300337}

\bibitem[{{Fern{\'a}ndez} {et~al.}(2016){Fern{\'a}ndez}, {Gim}, {van Gorkom}, {Yun}, {Momjian}, {Popping}, {Chomiuk}, {Hess}, {Hunt}, {Kreckel}, {Lucero}, {Maddox}, {Oosterloo}, {Pisano}, {Verheijen}, {Hales}, {Chung}, {Dodson}, {Golap}, {Gross}, {Henning}, {Hibbard}, {Jaff{\'e}}, {Donovan Meyer}, {Meyer}, {Sanchez-Barrantes}, {Schiminovich}, {Wicenec}, {Wilcots}, {Bershady}, {Scoville}, {Strader}, {Tremou}, {Salinas}, \& {Ch{\'a}vez}}]{2016ApJ...824L...1F}
{Fern{\'a}ndez}, X., {Gim}, H.~B., {van Gorkom}, J.~H., {et~al.} 2016, \apjl, 824, L1, \dodoi{10.3847/2041-8205/824/1/L1}

\bibitem[{{Freudling} {et~al.}(2011){Freudling}, {Staveley-Smith}, {Catinella}, {Minchin}, {Calabretta}, {Momjian}, {Zwaan}, {Meyer}, \& {O'Neil}}]{2011ApJ...727...40F}
{Freudling}, W., {Staveley-Smith}, L., {Catinella}, B., {et~al.} 2011, \apj, 727, 40, \dodoi{10.1088/0004-637X/727/1/40}

\bibitem[{{Giovanelli} {et~al.}(2005){Giovanelli}, {Haynes}, {Kent}, {Perillat}, {Saintonge}, {Brosch}, {Catinella}, {Hoffman}, {Stierwalt}, {Spekkens}, {Lerner}, {Masters}, {Momjian}, {Rosenberg}, {Springob}, {Boselli}, {Charmandaris}, {Darling}, {Davies}, {Garcia Lambas}, {Gavazzi}, {Giovanardi}, {Hardy}, {Hunt}, {Iovino}, {Karachentsev}, {Karachentseva}, {Koopmann}, {Marinoni}, {Minchin}, {Muller}, {Putman}, {Pantoja}, {Salzer}, {Scodeggio}, {Skillman}, {Solanes}, {Valotto}, {van Driel}, \& {van Zee}}]{2005AJ....130.2598G}
{Giovanelli}, R., {Haynes}, M.~P., {Kent}, B.~R., {et~al.} 2005, \aj, 130, 2598, \dodoi{10.1086/497431}

\bibitem[{{Gould} \& {Salpeter}(1963)}]{1963ApJ...138..393G}
{Gould}, R.~J., \& {Salpeter}, E.~E. 1963, \apj, 138, 393, \dodoi{10.1086/147654}

\bibitem[{{Haynes} {et~al.}(2011){Haynes}, {Giovanelli}, {Martin}, {Hess}, {Saintonge}, {Adams}, {Hallenbeck}, {Hoffman}, {Huang}, {Kent}, {Koopmann}, {Papastergis}, {Stierwalt}, {Balonek}, {Craig}, {Higdon}, {Kornreich}, {Miller}, {O'Donoghue}, {Olowin}, {Rosenberg}, {Spekkens}, {Troischt}, \& {Wilcots}}]{2011AJ....142..170H}
{Haynes}, M.~P., {Giovanelli}, R., {Martin}, A.~M., {et~al.} 2011, \aj, 142, 170, \dodoi{10.1088/0004-6256/142/5/170}

\bibitem[{{Hoppmann}(2014)}]{2014UWAThesisH}
{Hoppmann}, L. 2014, PhD thesis, University of Western Australia

\bibitem[{{Hoppmann} {et~al.}(2015){Hoppmann}, {Staveley-Smith}, {Freudling}, {Zwaan}, {Minchin}, \& {Calabretta}}]{2015MNRAS.452.3726H}
{Hoppmann}, L., {Staveley-Smith}, L., {Freudling}, W., {et~al.} 2015, \mnras, 452, 3726, \dodoi{10.1093/mnras/stv1084}

\bibitem[{{Jones} {et~al.}(2018){Jones}, {Haynes}, {Giovanelli}, \& {Moorman}}]{2018MNRAS.477....2J}
{Jones}, M.~G., {Haynes}, M.~P., {Giovanelli}, R., \& {Moorman}, C. 2018, \mnras, 477, 2, \dodoi{10.1093/mnras/sty521}

\bibitem[{{Jones} {et~al.}(2016){Jones}, {Papastergis}, {Haynes}, \& {Giovanelli}}]{2016MNRAS.457.4393J}
{Jones}, M.~G., {Papastergis}, E., {Haynes}, M.~P., \& {Giovanelli}, R. 2016, \mnras, 457, 4393, \dodoi{10.1093/mnras/stw263}

\bibitem[{{Kennicutt} \& {Evans}(2012)}]{2012ARA&A..50..531K}
{Kennicutt}, R.~C., \& {Evans}, N.~J. 2012, \araa, 50, 531, \dodoi{10.1146/annurev-astro-081811-125610}

\bibitem[{{Koribalski} {et~al.}(2020){Koribalski}, {Staveley-Smith}, {Westmeier}, {Serra}, {Spekkens}, {Wong}, {Lee-Waddell}, {Lagos}, {Obreschkow}, {Ryan-Weber}, {Zwaan}, {Kilborn}, {Bekiaris}, {Bekki}, {Bigiel}, {Boselli}, {Bosma}, {Catinella}, {Chauhan}, {Cluver}, {Colless}, {Courtois}, {Crain}, {de Blok}, {D{\'e}nes}, {Duffy}, {Elagali}, {Fluke}, {For}, {Heald}, {Henning}, {Hess}, {Holwerda}, {Howlett}, {Jarrett}, {Jones}, {Jones}, {J{\'o}zsa}, {Jurek}, {J{\"u}tte}, {Kamphuis}, {Karachentsev}, {Kerp}, {Kleiner}, {Kraan-Korteweg}, {L{\'o}pez-S{\'a}nchez}, {Madrid}, {Meyer}, {Mould}, {Murugeshan}, {Norris}, {Oh}, {Oosterloo}, {Popping}, {Putman}, {Reynolds}, {Rhee}, {Robotham}, {Ryder}, {Schr{\"o}der}, {Shao}, {Stevens}, {Taylor}, {van{\^A} der Hulst}, {Verdes-Montenegro}, {Wakker}, {Wang}, {Whiting}, {Winkel}, \& {Wolf}}]{2020Ap&SS.365..118K}
{Koribalski}, B.~S., {Staveley-Smith}, L., {Westmeier}, T., {et~al.} 2020, \apss, 365, 118, \dodoi{10.1007/s10509-020-03831-4}

\bibitem[{{Lang} {et~al.}(2003){Lang}, {Boyce}, {Kilborn}, {Minchin}, {Disney}, {Jordan}, {Grossi}, {Garcia}, {Freeman}, {Phillipps}, \& {Wright}}]{2003MNRAS.342..738L}
{Lang}, R.~H., {Boyce}, P.~J., {Kilborn}, V.~A., {et~al.} 2003, \mnras, 342, 738, \dodoi{10.1046/j.1365-8711.2003.06535.x}

\bibitem[{{Li} {et~al.}(2018){Li}, {Wang}, {Qian}, {Krco}, {Jiang}, {Yue}, {Jin}, {Zhu}, {Pan}, {Nan}, \& {Dunning}}]{2018IMMag..19..112L}
{Li}, D., {Wang}, P., {Qian}, L., {et~al.} 2018, IEEE Microwave Magazine, 19, 112, \dodoi{10.1109/MMM.2018.2802178}

\bibitem[{{Martin} {et~al.}(2010){Martin}, {Papastergis}, {Giovanelli}, {Haynes}, {Springob}, \& {Stierwalt}}]{2010ApJ...723.1359M}
{Martin}, A.~M., {Papastergis}, E., {Giovanelli}, R., {et~al.} 2010, \apj, 723, 1359, \dodoi{10.1088/0004-637X/723/2/1359}

\bibitem[{{Meyer} {et~al.}(2009){Meyer}, {Heald}, \& {Serra}}]{2009PRA...........M}
{Meyer}, M., {Heald}, G., \& {Serra}, P. 2009, Proceedings of Panoramic Radio Astronomy.
\newblock \url{https://pos.sissa.it/cgi-bin/reader/conf.cgi?confid=89}

\bibitem[{{Meyer} {et~al.}(2017){Meyer}, {Robotham}, {Obreschkow}, {Westmeier}, {Duffy}, \& {Staveley-Smith}}]{2017PASA...34...52M}
{Meyer}, M., {Robotham}, A., {Obreschkow}, D., {et~al.} 2017, \pasa, 34, 52, \dodoi{10.1017/pasa.2017.31}

\bibitem[{{Meyer} {et~al.}(2004){Meyer}, {Zwaan}, {Webster}, {Staveley-Smith}, {Ryan-Weber}, {Drinkwater}, {Barnes}, {Howlett}, {Kilborn}, {Stevens}, {Waugh}, {Pierce}, {Bhathal}, {de Blok}, {Disney}, {Ekers}, {Freeman}, {Garcia}, {Gibson}, {Harnett}, {Henning}, {Jerjen}, {Kesteven}, {Knezek}, {Koribalski}, {Mader}, {Marquarding}, {Minchin}, {O'Brien}, {Oosterloo}, {Price}, {Putman}, {Ryder}, {Sadler}, {Stewart}, {Stootman}, \& {Wright}}]{2004MNRAS.350.1195M}
{Meyer}, M.~J., {Zwaan}, M.~A., {Webster}, R.~L., {et~al.} 2004, \mnras, 350, 1195, \dodoi{10.1111/j.1365-2966.2004.07710.x}

\bibitem[{{Moorman} {et~al.}(2014){Moorman}, {Vogeley}, {Hoyle}, {Pan}, {Haynes}, \& {Giovanelli}}]{2014MNRAS.444.3559M}
{Moorman}, C.~M., {Vogeley}, M.~S., {Hoyle}, F., {et~al.} 2014, \mnras, 444, 3559, \dodoi{10.1093/mnras/stu1674}

\bibitem[{{Oman}(2022)}]{2022MNRAS.509.3268O}
{Oman}, K.~A. 2022, \mnras, 509, 3268, \dodoi{10.1093/mnras/stab3164}

\bibitem[{{Rauzy}(2001)}]{2001MNRAS.324...51R}
{Rauzy}, S. 2001, \mnras, 324, 51, \dodoi{10.1046/j.1365-8711.2001.04078.x}

\bibitem[{{Roberts} {et~al.}(2021){Roberts}, {Darling}, \& {Baker}}]{2021ApJ...911...38R}
{Roberts}, H., {Darling}, J., \& {Baker}, A.~J. 2021, \apj, 911, 38, \dodoi{10.3847/1538-4357/abe944}

\bibitem[{{Saintonge}(2007)}]{2007AJ....133.2087S}
{Saintonge}, A. 2007, \aj, 133, 2087, \dodoi{10.1086/513515}

\bibitem[{{Sancisi} {et~al.}(2008){Sancisi}, {Fraternali}, {Oosterloo}, \& {van der Hulst}}]{2008A&ARv..15..189S}
{Sancisi}, R., {Fraternali}, F., {Oosterloo}, T., \& {van der Hulst}, T. 2008, \aapr, 15, 189, \dodoi{10.1007/s00159-008-0010-0}

\bibitem[{{Spitzak} \& {Schneider}(1998)}]{1998ApJS..119..159S}
{Spitzak}, J.~G., \& {Schneider}, S.~E. 1998, \apjs, 119, 159, \dodoi{10.1086/313157}

\bibitem[{{Staveley-Smith} {et~al.}(2016){Staveley-Smith}, {Kraan-Korteweg}, {Schr{\"o}der}, {Henning}, {Koribalski}, {Stewart}, \& {Heald}}]{2016AJ....151...52S}
{Staveley-Smith}, L., {Kraan-Korteweg}, R.~C., {Schr{\"o}der}, A.~C., {et~al.} 2016, \aj, 151, 52, \dodoi{10.3847/0004-6256/151/3/52}

\bibitem[{{Staveley-Smith} {et~al.}(1996){Staveley-Smith}, {Wilson}, {Bird}, {Disney}, {Ekers}, {Freeman}, {Haynes}, {Sinclair}, {Vaile}, {Webster}, \& {Wright}}]{1996PASA...13..243S}
{Staveley-Smith}, L., {Wilson}, W.~E., {Bird}, T.~S., {et~al.} 1996, \pasa, 13, 243, \dodoi{10.1017/S1323358000020919}

\bibitem[{{Verheijen} {et~al.}(2007){Verheijen}, {van Gorkom}, {Szomoru}, {Dwarakanath}, {Poggianti}, \& {Schiminovich}}]{2007ApJ...668L...9V}
{Verheijen}, M., {van Gorkom}, J.~H., {Szomoru}, A., {et~al.} 2007, \apjl, 668, L9, \dodoi{10.1086/522621}

\bibitem[{{Westmeier} {et~al.}(2014){Westmeier}, {Jurek}, {Obreschkow}, {Koribalski}, \& {Staveley-Smith}}]{2014MNRAS.438.1176W}
{Westmeier}, T., {Jurek}, R., {Obreschkow}, D., {Koribalski}, B.~S., \& {Staveley-Smith}, L. 2014, \mnras, 438, 1176, \dodoi{10.1093/mnras/stt2266}

\bibitem[{{Wong} {et~al.}(2006){Wong}, {Ryan-Weber}, {Garcia-Appadoo}, {Webster}, {Staveley-Smith}, {Zwaan}, {Meyer}, {Barnes}, {Kilborn}, {Bhathal}, {de Blok}, {Disney}, {Doyle}, {Drinkwater}, {Ekers}, {Freeman}, {Gibson}, {Gurovich}, {Harnett}, {Henning}, {Jerjen}, {Kesteven}, {Knezek}, {Koribalski}, {Mader}, {Marquarding}, {Minchin}, {O'Brien}, {Putman}, {Ryder}, {Sadler}, {Stevens}, {Stewart}, {Stootman}, \& {Waugh}}]{2006MNRAS.371.1855W}
{Wong}, O.~I., {Ryan-Weber}, E.~V., {Garcia-Appadoo}, D.~A., {et~al.} 2006, \mnras, 371, 1855, \dodoi{10.1111/j.1365-2966.2006.10846.x}

\bibitem[{{Xi} {et~al.}(2022){Xi}, {Peng}, {Staveley-Smith}, {For}, \& {Liu}}]{2022PASA...39...19X}
{Xi}, H., {Peng}, B., {Staveley-Smith}, L., {For}, B.-Q., \& {Liu}, B. 2022, \pasa, 39, e019, \dodoi{10.1017/pasa.2022.16}

\bibitem[{{Xi} {et~al.}(2021){Xi}, {Staveley-Smith}, {For}, {Freudling}, {Zwaan}, {Hoppmann}, {Liang}, \& {Peng}}]{2021MNRAS.501.4550X}
{Xi}, H., {Staveley-Smith}, L., {For}, B.-Q., {et~al.} 2021, \mnras, 501, 4550, \dodoi{10.1093/mnras/staa3931}

\bibitem[{{Xi} {et~al.}(2024){Xi}, {Peng}, {Staveley-Smith}, {For}, {Liu}, {Chen}, {Yu}, {Ding}, {Guo}, {Zou}, {Xue}, {Wang}, {Brink}, {Zheng}, {Filippenko}, {Yang}, {Wei}, {Dai}, {Li}, {He}, {Jiang}, {Moiseev}, \& {Kotov}}]{2024ApJ...966L..36X}
{Xi}, H., {Peng}, B., {Staveley-Smith}, L., {et~al.} 2024, \apjl, 966, L36, \dodoi{10.3847/2041-8213/ad4357}

\bibitem[{{Zhang} {et~al.}(2024){Zhang}, {Zhu}, {Jiang}, {Cheng}, {Wang}, {Wang}, {Xu}, {Liu}, {Yu}, {Qian}, {Yu}, {Ai}, {Jing}, {Xu}, {Liu}, {Guan}, {Sun}, {Yang}, {Huang}, {Hao}, \& {FAST Collaboration}}]{2024SCPMA..6719511Z}
{Zhang}, C.-P., {Zhu}, M., {Jiang}, P., {et~al.} 2024, Science China Physics, Mechanics, and Astronomy, 67, 219511, \dodoi{10.1007/s11433-023-2219-7}

\bibitem[{{Zhang} {et~al.}(2019){Zhang}, {Wu}, {Li}, {Kr{\v{c}}o}, {Staveley-Smith}, {Tang}, {Qian}, {Liu}, {Jin}, {Yue}, {Zhu}, {Liu}, {Yu}, {Sun}, {Pan}, {Li}, {Gan}, \& {Yao}}]{2019SCPMA..6259506Z}
{Zhang}, K., {Wu}, J., {Li}, D., {et~al.} 2019, Science China Physics, Mechanics, and Astronomy, 62, 959506, \dodoi{10.1007/s11433-019-9383-y}

\bibitem[{{Zwaan} {et~al.}(1997){Zwaan}, {Briggs}, {Sprayberry}, \& {Sorar}}]{1997ApJ...490..173Z}
{Zwaan}, M.~A., {Briggs}, F.~H., {Sprayberry}, D., \& {Sorar}, E. 1997, \apj, 490, 173, \dodoi{10.1086/304872}

\bibitem[{{Zwaan} {et~al.}(2005){Zwaan}, {Meyer}, {Staveley-Smith}, \& {Webster}}]{2005MNRAS.359L..30Z}
{Zwaan}, M.~A., {Meyer}, M.~J., {Staveley-Smith}, L., \& {Webster}, R.~L. 2005, \mnras, 359, L30, \dodoi{10.1111/j.1745-3933.2005.00029.x}

\bibitem[{{Zwaan} {et~al.}(2004){Zwaan}, {Meyer}, {Webster}, {Staveley-Smith}, {Drinkwater}, {Barnes}, {Bhathal}, {de Blok}, {Disney}, {Ekers}, {Freeman}, {Garcia}, {Gibson}, {Harnett}, {Henning}, {Howlett}, {Jerjen}, {Kesteven}, {Kilborn}, {Knezek}, {Koribalski}, {Mader}, {Marquarding}, {Minchin}, {O'Brien}, {Oosterloo}, {Pierce}, {Price}, {Putman}, {Ryan-Weber}, {Ryder}, {Sadler}, {Stevens}, {Stewart}, {Stootman}, {Waugh}, \& {Wright}}]{2004MNRAS.350.1210Z}
{Zwaan}, M.~A., {Meyer}, M.~J., {Webster}, R.~L., {et~al.} 2004, \mnras, 350, 1210, \dodoi{10.1111/j.1365-2966.2004.07782.x}

\end{thebibliography}
\bibliographystyle{aasjournal}



\end{document}